# Population genomics of the *Wolbachia* endosymbiont in *Drosophila melanogaster*.

Mark F. Richardson[a,1], Lucy A. Weinert[b], John J. Welch[c], Raquel S. Linheiro[a], Michael M. Magwire[c,2], Francis M. Jiggins[c] & Casey M. Bergman[a,*]

[a]Faculty of Life Sciences, University of Manchester, Manchester, M13 9PT, UK
[b]UCL Genetics Institute, University College London, London, WC1E 6BT, UK
[c]Department of Genetics, University of Cambridge, Cambridge, CB2 3EH, UK

[1]Current Address: School of Life and Environmental Sciences,
Deakin University, Geelong, VIC 3217, Australia

[2]Current Address: Department of Genetics, North Carolina State University,
Raleigh, NC 27695, USA

*Address for correspondence:
Casey M. Bergman
Michael Smith Building
University of Manchester
Oxford Road
Manchester, M13 9PT
United Kingdom
Email: casey.bergman@manchester.ac.uk

6 Nov 2012

# Abstract

*Wolbachia* are maternally-inherited symbiotic bacteria commonly found in arthropods, which are able to manipulate the reproduction of their host in order to maximise their transmission. The evolutionary history of endosymbionts like *Wolbachia* can be revealed by integrating information on infection status in natural populations with patterns of sequence variation in *Wolbachia* and host mitochondrial genomes. Here we use whole genome resequencing data from 290 lines of *Drosophila melanogaster* from North America, Europe and Africa to predict *Wolbachia* infection status, estimate relative cytoplasmic genome copy number, and reconstruct *Wolbachia* and mitochondrial genome sequences. Overall, 63% of *Drosophila* strains were predicted to be infected with *Wolbachia* by our *in silico* analysis pipeline, which shows 99% concordance with infection status determined by diagnostic PCR. Complete *Wolbachia* and mitochondrial genomes show congruent phylogenies, consistent with strict vertical transmission through the maternal cytoplasm and imperfect transmission of *Wolbachia*. Bayesian phylogenetic analysis reveals that the most recent common ancestor of all *Wolbachia* and mitochondrial genomes in *D. melanogaster* dates to around 8,000 years ago. We find evidence for a recent global replacement of ancestral *Wolbachia* and mtDNA lineages, but our data suggest that the derived wMel lineage arose several thousand years ago, not in the 20th century as previously proposed. Our data also provide evidence that this global replacement event is incomplete, and is likely to be one of several similar incomplete replacement events that have occurred since the out-of-Africa migration that allowed *D. melanogaster* to colonize worldwide habitats. This study provides a complete genomic analysis of the evolutionary mode and temporal dynamics of the *D. melanogaster-Wolbachia* symbiosis, as well as important resources for further analyses of the impact of *Wolbachia* on host biology.

**Author Summary**

Host-microbe interactions play important roles in the physiology, development and ecology of many organisms. Studying how hosts and their microbial symbionts evolve together over time is crucial for understanding the impact that microbes have on host biology. With the advent of high-throughput sequencing technologies, it is now possible to obtain complete genomic information for hosts and their associated microbes. Here we use whole genome sequences from ~300 strains of the fruitfly *Drosophila melanogaster* to reveal the evolutionary history of this model species and its intracellular bacterial symbiont *Wolbachia*. The major findings of this study are that *Wolbachia* in *D. melanogaster* is inherited strictly through the egg with no evidence of horizontal transfer from other species, that the genealogies of *Wolbachia* and mitochondrial genomes are virtually the same, and that both *Wolbachia* and mitochondrial genomes show evidence for a recent incomplete global replacement event, which has left remnant lineages in North America, Europe and Africa. We also use the fact that *Wolbachia* and mitochondrial genomes have the same genealogy to estimate the rate of molecular evolution for *Wolbachia*, which allows us to put dates on key events in the history of this important host-microbe model system.

# Introduction

Heritable symbiotic associations are widespread and important components of animal physiology, development, ecology and evolution [1]. *Wolbachia* is a facultative endosymbiotic bacterium commonly found in association with insects and other arthropods (including spiders, scorpions, mites and terrestrial crustaceans), as well as filarial nematodes [2-4]. *Wolbachia* are α-Proteobacteria of the order Rickettsiales, a diverse group of bacterial species that exhibit commensal, mutualistic and parasitic relationships with their hosts [5,6]. Strains of *Wolbachia* can increase their frequency in natural populations by manipulating reproductive strategies of their hosts in various ways [5] and, in some species, by conferring resistance to certain viruses [7,8] or increased survivorship under nutritional stress [9]. *Wolbachia* are extremely widespread, with greater than 40% of arthropod species estimated to be infected [10-12], making it one of the most common endosymbionts known among all organisms.

Despite being primarily maternally transmitted through the cytoplasm of the egg, over long periods of evolutionary time *Wolbachia* strains are thought to undergo horizontal transfer among host species [13] and exchange sequences from different *Wolbachia* lineages by recombination [4,14]. However, the extent of these processes on microevolutionary timescales within host populations is less well understood, in part because of the limited resolution of the genetic markers used to study host and *Wolbachia* diversity [15]. The mode of transmission and other aspects of *Wolbachia* population dynamics can be examined indirectly by integrating information on *Wolbachia* infection status with patterns of host mitochondrial DNA (mtDNA) [16,17]. Since mitochondria and *Wolbachia* are thought to be maternally co-inherited, mtDNA variation can be used as a proxy to provide insights into the evolutionary history of *Wolbachia* in a host species such as the timing of an infection, its source, frequency and degree of spread through a population [18]. For example, if infected and uninfected individuals share mtDNA haplotypes, this indicates that there has either been imperfect maternal transmission or horizontal transfer of *Wolbachia* [19,20], while if they have different mtDNA haplotypes, it suggests maternal transmission rates are extremely high [18,21]. Similarly, if *Wolbachia* is strictly maternally transmitted, then strongly reduced levels of mtDNA variation may indicate that the bacterium has recently invaded the host population [18,22-24], while higher levels of mtDNA variation would suggest a more ancient association [25]. However, the fact that mtDNA is only a proxy for the *Wolbachia* genome itself often limits inferences that can be made from this important molecular marker.

Research on *Wolbachia* in *Drosophila melanogaster* and related species has made major contributions to our understanding of *Wolbachia* population dynamics [17,21,26,27]. The *Wolbachia* endosymbiont from *D. melanogaster* (wMel) was the first *Wolbachia* strain to have its genome fully sequenced [28]. Natural populations of *D. melanogaster* on all continents are known to be polymorphic for *Wolbachia* infection [1,17,19,29-32], but initial studies in *D. melanogaster* using marker loci revealed no sequence diversity in among *Wolbachia* isolates [17,33,34]. The availability of the complete sequence of *Wolbachia* from *D. melanogaster* permitted the identification of several structural variants that differentiate *Wolbachia* genotypes [35], opening up new possibilities to study the population genetics of *Wolbachia* in *D. melanogaster*. By genotyping these polymorphic markers in a panel of stocks isolated from nature at different time points, Riegler *et al.* [35] found evidence for a global replacement of a putatively ancestral *Wolbachia* genotype (called wMelCS) by a single genotype represented by the reference sequence (wMel) sometime in the 20$^{th}$ century. This scenario was reinforced by the work of Nunes *et al.* [19] who found that mtDNA haplotypes associated with the wMel strain had also increased in frequency during the late 20$^{th}$ century. However, neither of these studies used complete genomic information for either the *Wolbachia* or mtDNA and thus current inferences about the mode and temporal dynamics of *Wolbachia* and mtDNA in *D. melanogaster* remain incomplete.

Complete *Wolbachia* and mitochondrial genomes can be serendipitously assembled from the whole genome shotgun (WGS) sequences of host *Drosophila* species [36,37], and thus it is now possible to investigate the population dynamics of these cytoplasmic genomes using WGS data from population resequencing projects in *D. melanogaster* [38,39]. Here we use WGS sequence data from 290 strains of *D. melanogaster* to mine complete mitochondrial and (where present) *Wolbachia* genome sequences to study the co-ancestry of *Wolbachia* and *D. melanogaster*. Using these genomic resources, we reconstructed the genealogical history of the mitochondrial genome from all 290 strains, which revealed the presence of six major lineages. The pattern of *Wolbachia* infection across the mtDNA tree suggests that *Wolbachia* has been lost independently in many different populations through imperfect transmission. For the 179 infected strains, we show that the *Wolbachia* and mtDNA genealogies are fully congruent, suggesting that there was a single ancestral *Wolbachia* infection that has been vertically transmitted through the maternal cytoplasm without any paternal or horizontal transmission. The strict maternal transmission of *Wolbachia* and mtDNA in infected strains allows us to use estimates of the mtDNA mutation rate to calibrate the rates of evolution in *Wolbachia*. Using Bayesian phylogenetic analysis, we estimate dates for the major clades in the *Wolbachia* and mtDNA genealogies and show that most recent common ancestor of cytoplasmic lineages arose around 8,000 years ago. Using this same approach, we estimate that the rate of sequence evolution in *Wolbachia* is around 100-fold lower than the mutation rate at synonymous sites in mitochondrial genome and ten-fold lower than the mutation rate at noncoding sites in the nuclear genome of the host species. Patterns of *Wolbachia* and mtDNA molecular variation in a well-sampled North American population are inconsistent with a standard neutral model of molecular evolution, suggesting the action of natural selection or host population expansion acting on these genomes in the recent past. Finally, we present a biogeographic scenario for the recent evolution of *Wolbachia* and mtDNA that proposes multiple waves of incomplete replacement of pre-existing cytoplasmic lineages since *D. melanogaster* left Africa at the end of the Pleistocene, and discuss how the genomic resources reported here can be used to further evolutionary and functional analysis of insect-microbe symbioses.

## Results

**Shotgun sequencing of *D. melanogaster* accurately reveals *Wolbachia* infection status**

We used a reference-based short-read mapping pipeline to assemble the consensus sequence of complete mitochondrial and (where present) *Wolbachia* genomes in 290 strains of *D. melanogaster*: 174 strains from Raleigh, North Carolina provided by the *Drosophila* Genetic Reference Panel (DGRP) and 116 strains from a geographically diverse set of African (n=108) and European strains (n=8) provided by the *Drosophila* Population Genomics Project (DPGP). Detailed information about each strain can be found in Dataset S1. Overall, we analyzed 14,792,213,317 shotgun sequencing reads totalling 1,192,414,581,932 bp of DNA. We predicted a line to be infected with *Wolbachia* when a consensus sequence covered greater than 90% of the *Wolbachia* reference genome and the reference-based assembly had a mean depth of read coverage of greater than one. Both of these cutoffs were suggested by natural discontinuities in the distribution of coverage values in the data (Figure 1). Overall, 179 strains (61.8%) were predicted to be infected with *Wolbachia* by our *in silico* criteria (Table 1). The proportion of infected strains differs significantly in the DGRP (91/174; 52.2%) and DPGP samples (88/116; 75.9%) (Binomial test; $P < 1.4 \times 10^{-7}$).

We tested the validity of our *in silico* method for detecting *Wolbachia* infection using an experimental assay based on PCR amplification of the *Wolbachia wsp* gene in 167 of the 174 DGRP lines. Infection status predicted based on PCR was identical with our *in silico* predictions for 165/167 strains (98.8% concordance, Dataset S1). Only two lines gave different infection status predictions between procedures (DGRP38 and DGRP911). Both strains were scored as uninfected using *in silico* criteria, but scored as infected by PCR. Each of these lines have high overall WGS sequencing depth but very low depth and breadth of coverage in *Wolbachia* assemblies, suggesting that these discrepancies are not because of poor WGS sampling and may have instead arisen because of cross-contamination of fly stocks or DNA preparations for PCR, after preparation of WGS sequencing libraries. We note that in the DGRP sample, we did observe some strains that have a depth of coverage above one but only intermediate breadth of coverage, and that all of these that were tested by PCR were classified as uninfected. Assemblies of strains with intermediate coverage yield patches of sequence that differ in location across the *Wolbachia* genome from strain to strain, but which are all highly similar to other sequences from high coverage genomes. Moreover, intermediate coverage strains do not fall out as single clade on the mtDNA phylogeny as would be expected if they were vertically inherited in the cytoplasm (see below). Thus we do not believe intermediate coverage strains represent the presence of *Wolbachia* sequences in the nuclear genome segregating among DGRP strains, as has been observed in *D. ananassae* [40], or a secondary infection from another *Wolbachia* lineage. Rather we believe these data are consistent with low frequency polymorphic *Wolbachia* infections in the stocks used for sequencing that have subsequently been lost in culture, as would be predicted to occur occasionally under a mode of vertical inheritance with imperfect transmission. Despite the low rate of discrepancy and presence of some intermediate coverage strains, these results clearly demonstrate the high sensitivity and specificity of predicting *Wolbachia* infection in natural populations using whole-genome shotgun sequence of individual host strains.

**Estimates of relative mtDNA and *Wolbachia* copy number.**

The phenotypic effects of *Wolbachia* on insects often depend on the bacterial density in host cells [17,41]. Thus, we attempted to estimate the copy number of *Wolbachia* and mitochondrial genomes based on the depth of coverage for each assembly scaled relative to the nuclear genome to control for variation in overall WGS throughput (Figure 2). Relative copy number estimated in this manner represents an average across the tissue sampled, and obscures intra-individual variation across tissues. We found that the relative depth of coverage for both *Wolbachia* and mtDNA varied substantially among infected strains, but differed systematically between the DGRP strains (from diploid adult DNA) and DPGP strains (from haploid embryonic DNA), with a higher depth of coverage for both *Wolbachia* and mitochondrial DNA in the DGRP strains (Wilcoxon Rank Sum Tests; $P < 2.2 \times 10^{-16}$). This was also true for mtDNA from strains that were not infected with *Wolbachia* (Wilcoxon Rank Sum Tests; $P < 7.4 \times 10^{-9}$), and thus this pattern is not an artefact of the *Wolbachia* infection influencing the overall proportion of mtDNA or nuclear reads. For infected DGRP strains, the mean (standard deviation) *Wolbachia* coverage is on 5.57 (3.95) times greater than nuclear coverage, and mtDNA coverage is on average 32.9 (44.5) times greater than nuclear coverage. For infected DPGP strains, relative *Wolbachia* coverage is approximately the same as nuclear coverage (mean: 1.02; standard deviation: 1.84) and relative mtDNA coverage is 9.79 (24.7) times greater than nuclear coverage. Assuming nuclear coverage number represents a copy number of 2C in the DGRP strains (from diploid adults) and 1C in the DPGP strains (from haploid embryos), values for DPGP and half those for DGRP strains provide an estimate of cytoplasmic genome copy number relative to the haploid DNA content of the cell.

**Phylogenomics of *D. melanogaster* mtDNA in the Context of *Wolbachia* Infection**

To understand the relationship between *Wolbachia* infection status and mtDNA sequence variation, we reconstructed the recent genealogical history of the complete *D. melanogaster* mitochondrial genome using the entire set of 290 strains in the combined DGRP and DPGP sample (Figure 3). This analysis revealed six major intraspecific clades that we label I-VI. Clades I-IV also exhibit well-supported subclades within them, however we focus here on the higher-level aspects of the genealogy represented by these six major clades. Clade I contains the majority of the North America strains, one European strain, and multiple African strains in more basal locations. Clade II contains only African strains. Clade III contains mostly African strains, four European strains, nine North American strains, and the reference mtDNA from the dm3 *D. melanogaster* genome sequence. Clade IV is a small clade comprised of five strains only from Ethiopia. Clade V is only represented by two European strains. The most divergent clade (VI) contains two African strains and two North American strains. The composite mtDNA reference sequence NC_001709 also groups with this clade (see below).

*D. melanogaster* strains infected with *Wolbachia* are found across the entire mtDNA tree (Figure 3). All major clades have both infected and uninfected strains (with the exception of clade V, which comprises a small sample of uninfected strains). Assuming this pattern arose by a single infection with imperfect transmission, the distribution of infection status across the mtDNA genealogy provides indirect insight into the degree to which the *Wolbachia* infection has progressed. When a new *Wolbachia* strain invades it will tend to be associated with a single mtDNA type. As a new infection spreads, imperfect transmission of *Wolbachia* will cause mtDNA lineages originally associated with the infection also to be observed in uninfected flies. At equilibrium, the frequency of mtDNA haplotypes in infected and uninfected flies is ultimately expected to be the same [42]. Nunes *et al.* [19] previously found that mtDNA COI haplotypes in infected and uninfected strains differ significantly in *D. melanogaster,* which has been attributed to a recent change in the frequency of the *Wolbachia* infection. We tested for genetic differentiation between mtDNA sequences of infected and uninfected strains using a variant of $K_{st}$ [43] and found no evidence of subdivision (weighted mean $K_{st}$ across sampling locations: $K_{st}=0.01$, $P=0.88$), suggesting that the infection is not recent and that our sample does not contain mitochondrial lineages that predate the infection.

**Mitochondrial and *Wolbachia* genomes have congruent genealogies**

Strict maternal transmission of *Wolbachia* from a single infection would result in mtDNA and *Wolbachia* having congruent genealogies, while even low rates of paternal or horizontal transmission will break down this association. Rare paternal transmission has been observed previously in *D. simulans* laboratory crosses for both mtDNA [44] and *Wolbachia* [45]. Previous work by Nunes *et al.* [19] revealed a non-random association between *Wolbachia* genotypes and mtDNA COI haplotypes, but not a strict congruence. Such a pattern could arise from paternal or horizontal *Wolbachia* transmission among *D. melanogaster* mtDNA lineages or from using low-resolution molecular markers that obscure a true pattern of strict maternal inheritance. To test if *Wolbachia* is evolving under a strict maternal mode of transmission in *D. melanogaster*, we reconstructed independent genealogies of the *Wolbachia* and mitochondrial genomes for the 179 infected strains. Genealogies of both *Wolbachia* and mitochondria from infected strains reveal the presence of five distinct cytoplasmic lineages (clades I-IV and VI) in contemporary populations of *D. melanogaster*, with the majority of strains falling into clades I, II and III.

We found that genealogies of *Wolbachia* and mitochondria genomes reveal perfect congruence for all of the well-supported clades. In other words, no well-supported clade in the mitochondrial genealogy is contradicted, with strong support, in the *Wolbachia* genealogy and *vice versa* (Figure 4). We note that some discrepancies in perfect congruence between the ML topologies can be observed among weakly supported lineages at the tips of the tree, which arise from inherent uncertainties in genealogical inference among closely related lineages. To test the hypothesis that the *Wolbachia* and mitochondria genealogies are fully congruent, we used a Bayesian approach that assessed whether a model that allows separate topologies for each genome fits the data better than a model that forces a single topology on both genomes. The log marginal likelihood values (estimated from the harmonic mean) of the two models were -1,300,280.95 for the two-topology model and -1,300,271.06 for the single topology model. The log10 Bayes Factor of 4.30 shows decisive support for the single topology model, as did Akaike's information criterion (single topology: AIC value 2,600,722; two topology: 2,600,916) [46]. In contrast, when we randomly assigned *Wolbachia* to hosts within the same geographical location (i.e., *Wolbachia* strains were associated at random with hosts with same latitude and longitude as their true host), we found overwhelming support for the two-topology model (log10 Bayes Factor -944). Even greater support for two topologies was found when we randomly assigned *Wolbachia* to hosts within the same continent (Africa, Europe or USA; log10 Bayes factor -1215), or at random across the tree (log10 Bayes factor - 1539). Results from these randomised data sets suggest that the Bayes factor test would have had the power to detect incongruent genealogies, even if horizontal or paternal transmission was taking place solely within populations.

The support for a single genealogy describing both the *Wolbachia* and mitochondrial genomes from infected strains strongly implies a single ancestral infection followed by strict vertical transmission through the maternal cytoplasm, further supporting the inferences based on mtDNA genealogy of all strains above. Together with the fact that there is no reported evidence for horizontal transfer of mtDNA in *D. melanogaster* [17], the strong congruence of *Wolbachia* and mitochondria rules out the possibility that the divergent cytoplasmic clades IV, V and VI represent independent horizontal transfer events. Furthermore, a phylogeny of mtDNA from infected strains including an allele from the outgroup species *D. simulans* supports the placement of the root in the same position as is assumed by midpoint rooting without an outgroup (Dataset S2). Thus we conclude that maternal transmission with recurrent loss characterizes the mode of inheritance *Wolbachia* over the timescale of the divergence of sequences in this sample. Congruence of *Wolbachia* and mitochondrial genealogies also underscores the high quality of the genomic data and bioinformatics methods used here, since the biological signal of congruence could only be detected if the data have been processed accurately from WGS library preparation through to phylogenetic reconstruction. Furthermore, the strict maternal transmission implied by the congruence of *Wolbachia* and mitochondrial genealogies also argues against high rates of within-strain heteroplasmy for either *Wolbachia* or mtDNA. In fact, the proportion of individual read calls supporting the consensus sequence call at variable sites averaged across strains is very high for both *Wolbachia* (99.0%) and mtDNA (99.3%), suggesting that levels of heteroplasmy (or WGS sequence contamination) are very low in these data.

**Calibrating the timescale and rate of *Wolbachia* evolution**
Placing an absolute timescale on the evolution of bacterial symbionts allows patterns of bacterial evolution to be related to historical events in the evolution of their hosts. Likewise, insight into the evolutionary forces operating on a species can be substantially improved by understanding its spontaneous point mutation rate. Unfortunately, in *Wolbachia* there is no estimate of the spontaneous mutation rates that can be used to transform rates of sequence evolution into absolute time. We first attempted to measure the mutation rate directly in *Wolbachia* by applying our pipeline to *D. melanogaster* mutation accumulation lines from [47], but all were found to be uninfected (data not shown). Therefore we developed an indirect phylogenetic approach that assumes strict maternal transmission and incorporates prior information from empirically determined mutation rates in *D. melanogaster* mtDNA [48] to estimate rates of molecular evolution in the *Wolbachia* genome. Specifically, we concatenated *Wolbachia* and mtDNA sequences from the same infected strain and estimated a Bayesian dated phylogeny that placed a lognormal prior on mutation rates at mitochondrial third positions with mean -16.59613 and standard deviation 1/3 on the log scale. These values imply a mean rate of $6.2 \times 10^{-8}$ mutations per site per generation with 95% confidence intervals of approximately $3.6 \times 10^{-8}$ - $3.6 \times 10^{-7}$, and these values represent the average of the mutation rate estimates obtained from the Florida and Madrid lines studied in [48]. Estimates of dates in number of *D. melanogaster* generations were then converted into years by assuming ten generations per year (as in [19,49,50]). Our analysis also included a phylogeographic model, allowing us to infer the geographic locations of the ancestral strains simultaneously with the dated phylogeny. We note that our method estimates rates of sequence change along a coalescent tree, and strictly speaking these are neither mutation rates (because strongly deleterious mutations will not be observed) nor are they classical long-term neutral substitution rates (because they include slightly deleterious mutations that would not be observed in sequence divergence between species). Furthermore, we are only able to estimate the rate of sequence change in terms of host, not bacterial, generations. For these reasons, we use the term "short-term evolutionary rate" to describe the rate of sequence evolution estimated here from intraspecific variation, and measure changes in substitutions per site per host generation.

Using this approach, we find that the most recent common ancestor (MRCA) of all strains in the sample dates to approximately 8,000 years ago (ya) (Figure 5), substantially later than the estimated date of 16,000 ya for the migration event that allowed *D. melanogaster* to colonize non-African habitats [51]. The MRCA of clades I, II, III and IV dates to 5,000 ya, and is inferred to have arisen in Africa based on our phylogeographic model (see methods for details). The MRCA of the high frequency clades I, II and III dates to 2,200 years ago, and is also inferred to have an African origin. The appearance of the MRCA for the high-frequency clades I, II and III occurred within 1,000 years of this event (clade I MRCA: 1,800 ya; clade II MRCA: 1,400 ya; clade III MRCA: 1,200 ya). Subclades containing North America strains within clades I and III date to 700 and 375 ya, respectively (See Dataset S3 for details), prior to the estimated time of colonization of North American habitats in the 19[th] century [52,53].

These subclades also contain strains from Europe (Figure 3) and thus the MRCA of these subclades probably arose in Europe prior to their arrival in North America.

This analysis also provides estimates of the short-term molecular evolutionary rate for *Wolbachia* (Table 2), which we estimate from $3^{rd}$ codon positions to be $6.87 \times 10^{-10}$ substitutions/site/generation (95% Credible Interval: $2.88 \times 10^{-10}$ - $1.29 \times 10^{-9}$), roughly two orders of magnitude lower than the mutation rate in $3^{rd}$ codon positions of mtDNA (Table 2). The median *Wolbachia* substitution rate is also ten-fold lower than the *D. melanogaster* mutation rate of $3.5$-$5.8 \times 10^{-9}$ estimate from noncoding nuclear DNA [47,50]. In contrast to the *Drosophila* mtDNA, in the *Wolbachia* genome we find no evidence for differences in the estimated substitution rates for $1^{st}$ and $2^{nd}$ versus $3^{rd}$ codon positions (or noncoding DNA regions). We note that the estimates of *Wolbachia* substitution rates are in terms of *D. melanogaster* generations, not *Wolbachia* replications, and are thus directly comparable in molecular terms to the mutation rate estimates based on errors in replication of mitochondrial and nuclear DNA [47,48,50]. Tables of variable sites in the *Wolbachia* and mitochondrial genomes can be found in Dataset S4.

**Reconciliation of *Wolbachia* lineages based on genetic markers and whole genome data**

Previous work using genetic markers for both *Wolbachia* and mtDNA has suggested that there has been a global replacement of the *Wolbachia* strains in *D. melanogaster* during the $20^{th}$ century [19,35]. Riegler *et al.* [35] identified several different *Wolbachia* genotypes (wMel, wMel2, wMel3, wMelCS and wMelCS2) on the basis of a small number of structural variants including IS5 transposable elements, copy number variants and genome rearrangements. Subsequent work provided evidence that distinct mitochondrial lineages are associated with these *Wolbachia* genotypes [19,54]. Specifically, Ilinsky *et al.* [54] found strict association between two SNPs in the mtDNA COI gene and three wMel genotypes (wMel, wMelCS and wMelCS2) in an Ukrainian population, while Nunes *et al.* [19] found a non-random association between COI haplotypes and wMel genotypes in a worldwide sample.

To place our whole genome analyses in the context of this previous work, we identified the location of the *Wolbachia* IS5-family transposons (ISWpi1) insertions using an *in silico* transposable element mapping procedure related to that in [55]. Using this approach, we could discriminate between wMel-like (wMel or wMel2) and wMelCS-like (wMelCS or wMelCS2) genotypes for strains that had Illumina read lengths of greater than 75 bp. Based on this analysis, we infer that clades I-V are wMel-like genotypes and the basal clade VI is a wMelCS-like genotype. These inferences are supported by the placement of the *Wolbachia* reference genome (GenBank ID: AE017196), which defines the wMel genotype, in a sub-clade of North American strains within clade III (Figure 4). This placement of the wMel reference is consistent with its isolation from a *D. melanogaster* stock ($y,w^{67c23}$) that has its origin in North America [56]. We note that while direct confirmation that clade VI is the wMelCS genotype is in principle possible since this genome has been fully sequenced [57], we were not able to perform this analysis since these data have not been made publicly available at the time of writing.

In terms of mtDNA variants, we found that the diagnostic wMelCS-specific SNPs (T-2160/C-2187) of [54] were only present in clade VI, while all other strains had the diagnostic wMel-specific SNPs (C-2160/T-2187), further supporting the conclusion that our clades I-V are wMel-like. Likewise, we observe that haplotype 2, found by Nunes *et al.* [19] to be at highest frequency worldwide and preferentially associated with wMel-type *Wolbachia*, is present in clades I, III and IV (Figure 3). Clade II contains predominantly haplotypes 8 and 9, which are closely related to haplotype 2 and associated with wMel-type lineages [19]. Clade V, which is only represented by uninfected lineages in our sample, contains only haplotype 10, which is intermediate to haplotypes 1 and 2 but associated only with wMel-type lineages in [19]. In contrast, we found the rare haplotype 1, which is associated with wMelCS [19], to be present only in clade VI. Finally the phylogenetic placement of mtDNA reference sequences also supports the conclusion that clade VI represents the wMelCS lineages. The NC_001709 reference sequence, which is a composite of fragments from Oregon R and Canton S stocks [58-60], falls into clade VI, and complete mtDNA sequences from $w^{1118}$/Canton S (GenBank ID: FJ190105) and Oregon R (GenBank ID: AF200828) both carry the diagnostic wMelCS-specific 2160/C-2187 SNPs of [54] (results not shown). The placement of these Canton S like mtDNA sequences in clade VI is consistent with the fact that the wMelCS *Wolbachia* strain was derived from a Canton S strain [35]. Thus, all available evidence from *Wolbachia* and mitochondrial genomes support the inference that clades I-V represent wMel-like lineages and clade VI represent a wMelCS-like lineage.

**Evidence for non-neutral evolution of *Wolbachia* and mtDNA in North America**

A recent increase in frequency and spread of new cytoplasmic lineages through worldwide populations of *D. melanogaster* is expected to lead to low genetic diversity and an excess of low frequency polymorphisms. To test if these predictions are observed in the data, we estimated levels of nucleotide diversity and tested departures from a model of neutral equilibrium in the DGRP strains. We focused on the DGRP strains for this analysis since this project provides a large sample collected from the same location and time and thus fits the assumptions of the standard neutral model better than strains from the DPGP collection. We found low levels of nucleotide variation for both *Wolbachia* and mitochondrial genomes among DGRP lines (Table 3) relative to that found in the host nuclear genome ($\pi = 0.0056$ and $\theta = 0.0067$) [38]. Levels of mtDNA variation based on $\pi$ are somewhat lower than previous estimates based on marker loci (CytB, $\pi=0.0009$; ND5, $\pi=0.00149$; COI, $\pi=0.0018$) [23,61,62], however those based on $\theta$ are very similar to previous estimates (CytB, $\theta=0.0021$; ND5, $\theta=0.00298$) suggesting a deeper sampling of rare variants in our sample. Levels of *Wolbachia* nucleotide diversity are approximately an order of magnitude less than that observed for mtDNA. We also observe an excess of rare variants in both *Wolbachia* and mtDNA sequences relative to the expectations of the standard neutral model, with Tajima's D being significantly less than zero for all samples (Table 3). These results are consistent with a non-neutral process operating on *Wolbachia* and mitochondrial genomes in the DGRP population. This signal could result from the action of a recent selective sweep driving the global replacement of the wMel-like *Wolbachia* and mitochondrial lineages, with possible fitness effects arising from differential longevity [63], protection against viruses [7,8], or co-adaptation with the host [64]. Alternatively, the excess of rare variants may be explained by purifying selection on weakly deleterious mutations as has been proposed previously for mtDNA in *D. melanogaster* [23]. Similar patterns could also have been generated by demographic effects such a population size expansion of the host after *D. melanogaster* colonized non-African habitats [49].

**Discussion**

Using high-throughput shotgun sequencing data from several hundred strains of *D. melanogaster*, we have reconstructed complete genome sequences of the *Wolbachia* endosymbiont and mtDNA to study the recent evolutionary dynamics of these two important model organisms. We use these new genomic resources to estimate copy number of *Wolbachia* and mitochondrial genomes in the host cell, and to compare patterns of *Wolbachia* infection across the *D. melanogaster* mtDNA genealogy. We identify several distinct cytoplasmic lineages that show strong congruence between the *Wolbachia* and mtDNA genealogies. Our data support a single ancestral *Wolbachia* infection that has been inherited strictly by vertical transmission in the maternal cytoplasm. This observation allows us to use empirically determined rates of mtDNA evolution to calibrate rates of *Wolbachia* evolution. We show that the most recent common ancestor of the current *Wolbachia* infection in *D. melanogaster* dates to less than 10,000 years ago, and that patterns of molecular variation for the *Wolbachia* and mitochondrial genomes in a well-sampled North American population are inconsistent with a standard neutral population genetic model.

Our use of reference-based endosymbiont genome reconstruction from host whole genome shotgun sequences extends previous efforts to identify endosymbiont genomes on the basis of *de novo* assembly [36,65]. Our results also show that it is not necessary to purify *Wolbachia* prior to whole-genome shotgun sequencing (e.g [57]) in order to study the genetics and evolution of this microorganism in *D. melanogaster*. Furthermore, by demonstrating that host whole-genome shotgun sequences can accurately predict *Wolbachia* infection status, our work also shows that it is no longer necessary to rely only on indirect mtDNA-based analyses (e.g. [17]) or low-resolution techniques like diagnostic PCR or marker-based genotyping methods (e.g. [19,35]) in order to study *Wolbachia*, since the ultimate level of genetic resolution – complete *Wolbachia* genomic sequences – can now be achieved by a relatively easy and scalable protocol. Finally, we show that complete *Wolbachia* and mitochondrial genomes can be readily obtained from shotgun sequencing libraries of both adult and embryonic DNA in *D. melanogaster*, however our results would suggest that sampling from adults gives higher yields of cytoplasmic genomes (see below). Furthermore, given the maternal transmission of both mtDNA and *Wolbachia*, it may not be necessary to use the haploid embryonic DNA preparation technique of Langley *et al.* [66] to sample cytoplasmic genomes using WGS.

WGS provides an opportunity to study differences in copy number of various genomic regions in a library of sequences. By normalizing to a standard nuclear reference sequence (in order to control for variation in overall sequencing throughput) we obtained estimates of the abundance of *Wolbachia* and mitochondrial genomes in each strain relative to the *D. melanogaster* nuclear genome (Figure 2). This analysis revealed a greater abundance of mtDNA relative to *Wolbachia* across both the DGRP and DPGP strains, as well as a greater abundance of both cytoplasmic genomes in DGRP strains relative DPGP strains. The higher abundance of mtDNA is likely to reflect a real biological difference in the relative copy number of these cytoplasmic genomes since this trend is observed in both projects and could arise from a higher titre of mitochondria per cell or because mtDNA is multicopy in a given mitochondria. However, since DGRP strains were prepared from mixed sex adult flies [38] and DPGP samples were prepared from gynogenetic haploid embryos that have undergone whole-genome amplification [66], differences in relative abundance between these projects could arise either from (i) lower abundance of cytoplasmic DNA in embryos relative to adults, (ii) biases generated by the whole-genome amplification process that skews the ratio of cytoplasmic to nuclear DNA, or (iii) differences in copy number across populations. It is unwarranted to conclude that these observations reflect real geographic variation among populations until differences in sample preparation are excluded. Moreover, if relative abundance correlates with levels of infection or transmission rates we would expect the DPGP sample to have lower infection rates, which is opposite to what is observed (Table 1). Assuming that the differences between the DGRP and DPGP represent real biological differences between the adult and embryonic stages of the life cycle rather than technical artefacts, our results would suggest a strong reduction in relative cytoplasmic genome copy number during oogenesis or embryogenesis. Recent work using quantitative PCR (qPCR) has shown that relative copy number of mtDNA in adults is on the order of 200 copies per nuclear genome [67], suggesting that studying relative mtDNA copy number using a WGS approach provides lower estimates than qPCR. While this and other caveats prevent the straightforward interpretation of relative abundance in terms of actual copy number in the cell, reduced bacterial titre during oogenesis or embryogenesis would provide a simple stochastic mechanism [41] to explain the relatively high rate of imperfect transmission for *Wolbachia* in *D. melanogaster* [68].

Two lines of evidence presented here support the inference that the current *Wolbachia* infection in *D. melanogaster* arose once in the past and has been inherited by strict vertical transmission in the maternal cytoplasm, with subsequent loss of the infection in multiple populations worldwide. First, we observe that the *Wolbachia* infection is found in all major clades across the mtDNA genealogy (Figure 3), which is most parsimoniously interpreted in terms of a single gain and multiple losses, given that imperfect maternal transmission occurs at a high frequency in the wild [68]. While this conclusion has been made in the past based on mtDNA marker genes [17], the inference of an infection across the deepest node in a mtDNA tree itself does not exclude the possibility of more than one infection by horizontal transfer. Because maintenance of a *Wolbachia* infection in the face of imperfect transmission implies some form of positive transmission bias increasing frequency of infected lineages, uninfected mtDNA lineages predating an infection will be rapidly lost [69]. Thus, we only expect to see mtDNA lineages related to those of infected strains in nature, even if there were multiple independent *Wolbachia* infections. Vertical transmission is secondly supported by the strong congruence between *Wolbachia* and mtDNA genealogies in infected strains (Figure 4). Intraspecific genealogical congruence is only consistent with strict vertical co-transmission of both cytoplasmic genomes [70], but not with horizontal transfer of *Wolbachia* from another species on a vertically evolving mtDNA lineage. In principle, the pattern of co-transmission we observe could occur through simultaneous introgression of both *Wolbachia* and mtDNA lineages from a sister species through hybridization. However there is no evidence for introgression of mtDNA from sister species into *D. melanogaster* [17,61]. Taken together, the phylogenetic evidence strongly supports a single infection with vertical transmission through the maternal cytoplasm at least as far back as the time to the MRCA of the sample.

Our conclusion of a single infection is consistent with previous genetic evidence that *Wolbachia* gives rise to a single cytoplasmic incompatibility type in *D. melanogaster* [17]. Our inference of strict maternal transmission contrasts with previous reports for rare paternal transmission of *Wolbachia* in *D. simulans* under laboratory conditions [45,71]. However, our conclusion that paternal transmission in nature occurs rarely, if ever, is supported directly by experimental evidence in *D. melanogaster* [68] and indirectly by analysis of mtDNA frequencies in *D. simulans* [21]. Two consequences of strict maternal transmission are that heteroplasmy would be expected to be rare or non-existent in nature, and that no paternal lineages would be present in an individual for homologous recombination to occur. Finally, while interspecific horizontal transfer of *Wolbachia* may occur on large evolutionary timescales [5] and has been inferred to occur within species of some arthropods (e.g. [72]), we find no evidence of horizontal transfer within current *D. melanogaster* populations, at least within the limits of detection afforded by essentially complete genomic coverage and deep population sampling. However, we cannot reject some horizontal transfer between individuals that have identical or nearly identical sequences using a phylogenetic approach.

Our observation that genealogies for *Wolbachia* and its host *D. melanogaster* are congruent over short evolutionary timescales (Figure 4) is also important because it allows us to use information from the host species to calibrate rates of sequence evolution for a bacterial species that lacks a fossil record [73,74]. To date, this approach has not been applied to *Wolbachia* because of evidence for horizontal transfer over longer evolutionary time periods [5]. Using a novel Bayesian approach to calibrate *Wolbachia* evolutionary rates using empirically determined mtDNA mutation rates from *D. melanogaster* [48], we find that the evolutionary rate for all classes of sites studied in *Wolbachia* are a 100-fold lower than silent sites in host mtDNA (Table 2) and ten-fold lower than noncoding sites in host nuclear DNA [47,50]. Moreover, in contrast to the five-fold reduction in mutation rates observed for coding sites relative to silent sites in mtDNA, we find no difference in the short-term evolutionary rate for coding and silent sites in *Wolbachia*. Assuming that changes at silent sites are selectively unconstrained, this observation suggests that our estimates of the rate of short-term sequence evolution for both silent and coding sites are closer to the mutation rate than the long-term neutral substitution rate. This pattern of molecular evolution is consistent with a mode of purifying selection operating on *Wolbachia* protein sequences that is either relaxed or has had insufficient time to purge newly arising slightly deleterious mutations from the population [75]. We favor the interpretation of the delayed action of purifying selection since sequence divergence in *Wolbachia* genes between *D. melanogaster* and *D. simulans* suggests evidence for purifying selection (median $K_a$=6.2x10$^{-3}$; median $K_s$=3.2x10$^{-2}$ [6]). Our estimate of the evolutionary rate for *Wolbachia* is more than 30-fold lower than the short-term evolutionary rate at silent sites estimated for the *Buchnera* endosymbiont of aphids (2.2x10$^{-8}$ substitutions/site/host generation [76] assuming 10 aphid generations per year [77]). Moran *et al.* [76] also found than silent sites in *Buchnera* had a two-fold higher short-term evolutionary rate relative to the genome-wide estimate that includes coding sites, in contrast to what we observe in *Wolbachia*. The principal observation that *Wolbachia* has a much lower short-term evolutionary rate than *Buchnera* is consistent with *Wolbachia* having functional DNA repair pathways [28], and helps explains why *Wolbachia* has not undergone such extensive genome erosion as is observed in *Buchnera* [78]. Moreover, the different rate and pattern of mutation between *Buchnera* and *Wolbachia* genomes argues against the application of single universal evolutionary model for studying molecular divergence among bacterial endosymbiotic lineages [74].

Our work also provides important insights into the debate about the recent biogeographic history of *Wolbachia* in *D. melanogaster*. Solignac *et al.* [17] first proposed that the *Wolbachia* infection in *D. melanogaster* arose once and has been inherited maternally with subsequent loss by imperfect transmission, consistent with the findings presented here. These authors suggested that the infection arose sometime after the split of *D. melanogaster* from its sister species *D. simulans* but prior to the MRCA of all *D. melanogaster* mtDNA sequences, which they estimate to be around 500 thousand ya [17]. We estimate the date for the MRCA for mtDNA to be much younger at around 8,000 years, which provides a minimum bound on the age of the infection. However, because of this very recent coalescence event, we cannot say much about the age of the infection prior to this time. Riegler *et al.* [35] postulated that after this initial infection, a wMel-like lineage arose sometime in the late 19$^{th}$ or early 20$^{th}$ century in North America, which replaced ancestral wMelCS-like lineages on all continents in the 20$^{th}$ century. This scenario of a recent global replacement of wMelCS-like by wMel-like lineages was supported by the work of Nunes *et al.* [19], who showed that mtDNA haplotypes preferentially associated with wMel-like lineages have become more prevalent in the late 20$^{th}$ century. Assuming our inferences that cytoplasmic clades I-V represent wMel-like lineages and clade VI represent a wMelCS-like lineage are correct, the widespread geographic distribution of strains in clades I-V (wMel-like) and the basal location of clade VI (wMelCS-like) is consistent with the Riegler *et al.* [35] global replacement hypothesis. However, our data and others [19,31] demonstrate that wMelCS-like lineages still persist naturally at low frequency in North American and Eurasian populations in the 21$^{st}$ century, and thus this replacement event is clearly incomplete. Furthermore, our inference of the date and geographic location of the ancestor of wMel-like lineages is inconsistent with an origin in North America in the late 19$^{th}$ or early 20$^{th}$ century. Rather, we find that the MRCA of the wMel-like lineages arose several thousand years ago, long enough ago to allow subsequent diversification into distinct clades. The fact that the MRCA of the wMel-like subclades present in North America date to greater than 300 ya provides further evidence for the inference that the replacement of wMelCS-like lineages occurred prior to colonization of North America by *D. melanogaster* in the late 19$^{th}$ century [52,53]. We propose instead that the wMel-like replacement event occurred in the Old World and was incomplete, leaving remnant wMelCS-like and basal wMel-like cytoplasmic lineages in the Afrotropical and Palearctic regions. Sampling of both high frequency wMel-like and low frequency wMelCS-like lineages from these regions during colonization of North America would have led to the mixture of cytoplasmic lineages currently observed in the DGRP sample. The possibility of some populations harbouring remnant cytoplasmic lineages, together with the observation of populations that are entirely free of the *Wolbachia* infection [17,31], suggests substantial geographic structure with respect to the *Wolbachia* infection in *D. melanogaster*. The clearest evidence for this model comes from the clade IV lineage being observed only in the Ethiopian sample from Dodola (ED), which also shows the highest genetic differentiation in the nuclear genome among all DPGP populations [39].

Finally, the resources and approaches presented here offer the possibility for wider application in evolutionary and functional genomics. For example, it is now possible to use the presence or absence of *Wolbachia* infection as a control factor in genome-wide association studies of host traits based on the DGRP lines [38]. In fact it is also now possible to treat *Wolbachia* presence/absence or the relative abundance of *Wolbachia* as traits in genome-wide association studies to identify variation in host genes involved in the modulation of *Wolbachia* infection. Further studies could investigate the relative rates of various types of mutation observed in the *Wolbachia* genome, including insertions/deletion and larger structural variants not studied here, or attempt to identify the functional effects of variants in the *Wolbachia* genome that might underpin traits such as the increased virulence of strains such as the wMelPop "popcorn" strain [79]. Reference-based shotgun sequence assembly using the DGRP and DPGP lines could be also applied to other endosymbiotic and non-endosymbiotic bacteria that are known to be associated with *D. melanogaster* [80]. For example, a preliminary analysis of *Spiroplasma* infection in the DGRP and DPGP lines using marker loci as references (GenBank ID: FJ657061, FJ657121, FJ657249) revealed that only two closely related strains from Uganda (UM37 & UM526) are likely to be infected with this endosymbiont, both of which are also infected with *Wolbachia*. This result is consistent with the only other report of *Spiroplasma* infection in African populations of *D. melanogaster* being located in Uganda [81], and the common occurrence of strains that are co-infected with these two endosymbionts [82]. In the absence of a complete *D. melanogaster Spiroplasma* reference sequence, this result motivates an attempt to reconstruct the *Spiroplasma* genome from these DPGP strains using a *de novo* assembly technique [36,65]. With continuing advances in high-throughput sequencing and the proof of principle presented here, it is now feasible to consider the comprehensive co-evolutionary analysis of symbionts and their hosts in a population genomics context.

## Materials and Methods

### Data sources

WGS sequences of *D. melanogaster* strains were downloaded from the NCBI Short Read Archive (SRA) from two projects: (i) 176 inbred lines sequenced by the *Drosophila* Genetic Reference Panel (DGRP; NCBI SRA project: SRP000694) from a single population in Raleigh, NC, USA [38]; and (ii) 118 "core" isofemale lines sequenced by the *Drosophila* Population Genomics Project (DPGP; NCBI SRA project: SRP005599) from multiple populations in Africa and a single population in Europe [39]. Two DGRP strains (SRS003443 and SRS003448) that previously have been proposed to contain chimeric Illumina reads [55] were excluded from this analysis. Sequences from strains with multiple SRA sequencing run accessions were concatenated into single fastq files before further processing. Statistics of total read count and total sequence length per strain were calculated using seqtk (https://github.com/lh3/seqtk) and faSize (http://genome.ucsc.edu/admin/git.html).

### Reference-based mitochondrial and *Wolbachia* genome assembly

Fastq sequences were mapped to reference genomes with BWA version 0.5.9-r16 [83] using default parameters and converted to BAM format with SAMtools version 0.1.16 [84]. Reads were mapped to three reference sequences: (i) a mitochondrial reference sequence extracted from the Release 5 genome sequence (chrU:5288528-5305749); (ii) the *D. melanogaster Wolbachia* endosymbiont reference genome (GenBank ID: AE017196); and (iii) an equivalently-sized (1.2 Mb) nuclear region randomly chosen from the middle of *D. melanogaster* chromosome arm 3L (GenBank ID: NT037436; positions 10000000-11200000). We used the chrU version of the mtDNA sequence as our reference since it represents the true mtDNA sequence from the *D. melanogaster* $y^1$, $cn^1$, $bw^1$, $sp^1$ strain [85], not the composite mtDNA sequence provided with the Release 5 genome sequence (GenBank ID: NC_001709). Since not all genomes or runs had paired-end data available, reads were mapped in single-ended mode for consistency. Statistics of total mapped read count and total sequence length per strain and reference file were calculated using SAMtools to ensure that all reads from the input fastq were accounted for in the BAM files.

### Variant calling and consensus sequence generation

Variant base calling followed a standard SAMtools version 0.1.16 pileup pipeline [84]. Individual strain consensus fastq sequences were generated where minimum and maximum read depths were set to 10 and 100, respectively, using pileup2fq.pl, and converted to fasta using a custom PERL script. Insertions relative to the reference sequence were excluded and deletions relative to the reference sequence were coded as N's. Where necessary, individual consensus sequences were extended by adding N's to the 3' end or deleting nucleotides from the 3' end to produce consensus sequences with the same length as the reference sequence.

**Prediction of *Wolbachia* infection status from WGS data**

*Wolbachia* infection status was determined automatically by calculating the mean depth of coverage of the assembly and breadth of coverage of the consensus sequences. Depth of coverage at each nucleotide in the reference sequence was estimated from BAM files using the genomeCoverageBed utility from BEDtools version 2 [86], and mean depth of coverage was calculated from BEDtools output by a custom PERL script. Breadth of coverage is defined as the proportion of nucleotides with non-N base calls in the consensus sequences and was calculated using a custom BioPerl-enabled PERL script [87]. A line was scored as "infected" when breadth of coverage was greater than 90% of the *Wolbachia* genome and mean depth of coverage was greater than one. Conversely, when a consensus sequence covered less than 90% of the *Wolbachia* genome and mean coverage was less than one, a line was scored as "uninfected." Infection status and other metadata for each strain can be found in Dataset S1.

**Experimental determination of *Wolbachia* infection status**

DNA was extracted from pools of approximately 20 individuals from each DGRP fly line using a method based on Chelex 100 resin [88]. We used a diagnostic PCR to test for the presence of the *Wolbachia wsp* gene using the primers wsp81F (5'-tgg tcc aat aag tga tga aga aac-3') and wsp691r (5'-aaa aat taa acg cta ctc ca-3'). The conditions for this diagnostic reaction were 35 cycles of 94ºC for 15 seconds, 55ºC for 30 seconds and 72ºC for 1 minute. Each *Wolbachia* PCR was repeated twice to check that the results were consistent. The success of DNA extraction was confirmed using the CHK_F-CHK_R and *Doc*1420_F-CHK_R primer pairs that amplify *Drosophila* nuclear genomic DNA [89]. The conditions for this control reaction were 35 cycles of 94ºC for 15 seconds, 55ºC for 30 seconds and 72ºC for 30 seconds).

**Phylogenomic analysis**

Multiple alignments were constructed simply by concatenating individual reference-based fasta consensus sequence files. For these analyses, we also included mitochondrial and *Wolbachia* reference sequences to place them in the context of global sequence diversity. All alignment columns that had an N in any strain (which can represent either a fully ambiguous character or a deletion relative to the reference) were then removed. The resulting "essentially complete" multiple sequence alignments were then converted to Phylip format using Seqret (http://emboss.sourceforge.net/) and used to reconstruct phylogenies with RAxML version 7.0.4 [90]. Maximum likelihood tree searches were conducted using a general time reversible (GTR) model of nucleotide substitution with Γ rate heterogeneity, with all model parameters estimated by RAxML. Trees were inferred using a combined approach, with an initial 100 bootstrap replicates and a full ML search for the best-scoring tree, using the rapid bootstrap algorithm [91]. The best-scoring ML trees were visualized and annotated in FigTree version 1.3.1 (http://tree.bio.ed.ac.uk/software/figtree). Strain identifiers in major clades were selected using Hypertree version 1.2.2 [92]. Bootstrap maximum likelihood trees in Newick format can be found in Dataset S2.

### *In silico* genotyping of IS5 elements in *Wolbachia* assemblies

We predicted the presence or absence of IS5 transposons in *Wolbachia* genomes at two loci (WD0516/7 and WD1310) defined by [35] to be diagnostic for *Wolbachia* genotypes. To do this we used a BLAT-based mapping strategy to identify transposable element flanking sequences similar to that reported in [55]. The presence of an IS5 insertion site at WD0516/7 (AE017196:507322-509812) and absence of an IS5 at WD1310 (AE017196:1251363-1252108) indicated a wMel-type *Wolbachia* strain (wMel or wMel2), while the converse configuration indicated a wMelCS-type strain (wMelCS or wMelCS2). If IS5 is absent from both loci, then the infecting strain is predicted to be the wMel3 genotype [35]. Using this procedure, we were able to identify both wMel-type and wMelCS-type strains in our data, but the rare wMel3 was never observed.

### COI haplotype analysis

We extracted sequences corresponding to the cytochrome c oxidase subunit I (COI) gene (ChrU: 5,290,184-5,290,738) from our mtDNA assemblies and merged them into a previously reported multiple alignment of COI sequences for numerous global populations [19]. Haplotype analysis was then conducted on the resulting alignment in DnaSP version 5 [93].

Sequences in our dataset that were found to be in previously identified haplotypes were given haplotype designations according to Nunes *et al.* [19]. New haplotypes were numbered sequentially starting from 20, after the highest haplotype number (19) identified in [19].

**Bayesian analysis of dates and rates of *Wolbachia* evolution**

To obtain a dated genealogy from the *Wolbachia* multiple alignment, we performed a Bayesian molecular evolutionary analysis using BEAST v.1.7.1 [94,95] incorporating phylogeographic data [96-98]. This analysis assumed a three-region phylogeographic model, with each strain labelled as coming from Africa, Europe or North America. We modelled the locations of the ancestral strains over the tree using a continuous-time Markov chain of these three locations, with a non-reversible infinite rate matrix parameterizing the transitions between them, and therefore allowing the rate of transmission from Africa to Europe to differ from the rate of transmission from Europe to Africa, for example [96,97]. We chose a constant-population-size coalescent prior for the relative node heights, and all other priors were set at their default values as assigned by the software BEAUti [95]. We note that using an exponentially-changing population-size coalescent prior for the relative node heights made no qualitative difference to our results. To check convergence of the posterior distributions of model parameters, we ran two independent MCMC chains, visualizing results in the program Tracer v1.5, before discarding burn-in as appropriate.

For Bayesian phylogenetic analyses, we used an annotated version of the multiple alignments of *Wolbachia* and mtDNA sequences from infected strains (omitting reference sequences) that labelled all alignment columns as one of 5 classes of site: $1^{st}$ codon position, $2^{nd}$ codon position, $3^{rd}$ codon position, noncoding RNA gene, or noncoding DNA intergenic region. We excluded all sites with gaps or fully ambiguous characters in any strain from this annotated alignment (as above) and then concatenated *Wolbachia* and mitochondrial genomes into a single combined sequence from each infected strain. We next performed an initial Bayesian analysis to identify classes that showed poor convergence because of low number of informative sites, which were then removed from the analysis. The final dataset included five rate categories: (i) mitochondrial first and second codon positions, (ii) mitochondrial third positions, (iii) *Wolbachia* first and second codon positions, (iv) *Wolbachia* third codon positions, and (v) *Wolbachia* noncoding DNA. Following [99], each of these five rate categories was assigned its own HKY+Γ model of molecular evolution. BEAST XML files and maximum clade consensus trees in Newick format can be found in Dataset S3.

**Test of genealogical congruence**

To determine whether *Wolbachia* and mitochondrial genomes have the same evolutionary history, we implemented a Bayesian incongruence test similar to that of [100] using BEAST v.1.7.1. Specifically, we compared the fit to our data of one- and two-topology models, first with the *Wolbachia* and mitochondrial alignments constrained to have the same dated topology, then allowing *Wolbachia* and mitochondrial partitions to have distinct topologies and node ages. All models and partitions were as described above, with the exception of the prior temporal information, which applied to mitochondrial sites only, and so provided no constraints on the *Wolbachia* data in the two-topology analysis. Accordingly, for the congruence test we set a prior on the root age that applied to all partitions. We chose this prior to correspond loosely to the results of the full dating analysis, and so specified a normal distribution of mean 76,287 and standard deviation 10. To compare the fit of the one- and two-topology models, we computed Bayes Factor values from the difference in log marginal likelihood values estimated using the harmonic mean of the log likelihoods using Tracer v1.5 and calculated Akaike's information criterion through MCMC values using the method described in [46]. BEAST XML files for the congruence test can be found in Dataset S3.

**Levels and patterns of mitochondrial and *Wolbachia* nucleotide diversity**

Levels of polymorphism for both *Wolbachia* and mtDNA were estimated as $\pi$ [101] and $\theta_w$ [102] based on the total number of mutations using Variscan version 2.0.2 [103]. To test if the frequency spectrum of polymorphisms conformed to predictions of the standard neutral model of molecular evolution, we calculated Tajima's *D* [104] using Variscan from the DGRP sample. Significance levels were based on 10,000 coalescent simulations using ms [105] assuming no recombination and conditioned on the number of variable sites [106]. Tables of variable sites from alignments with indels and fully ambiguous sites removed can be found in Dataset S4.

To test whether there is genetic subdivision in the mtDNA of infected and uninfected flies within a population, we used the mitochondrial genomes to calculated a variant of the $K_{st}$ statistic [43] between infected and uninfected flies. The genetic distance between mitochondrial genomes was calculated as the patristic distance on the mtDNA genealogy (inferred using a strict molecular clock in BEAST as described above for the concatenated alignment, but removing *Wolbachia* partitions), and $K_{st}$ was calculated between the infected and uninfected flies within each population. We assessed the significance of this statistic by permuting the infection status of the flies within each population (defined as samples collected from the same latitude and longitude).

## Acknowledgments

We thank Daniela Nunes for providing mtDNA COI data files and alerting us to the differences among mtDNA reference sequences, Marc Suchard, Guy Baele and Philippe Lemey for helpful comments on the Bayesian phylogenetic analysis, Douda Bensasson for Pileup file parsing code, and Anna Page and Florence Gutzwiller for assistance with the analysis of intermediate coverage DGRP strains. We thank Daniela Nunes, Christian Schlotterer, Danny Rozen, Douda Bensasson, Imran Rahman, John Pool, Yury Ilinsky, Iryna Kozeretska, three anonymous reviewers and members of the Bergman Lab for helpful discussion and comments on the manuscript. We are especially indebted to Trudy Mackay, Stephen Richards, John Pool and Chuck Langley for pre-publication access to the DGRP and DPGP datasets and for their encouragement throughout this project.

# References

1. Mateos M, Castrezana SJ, Nankivell BJ, Estes AM, Markow TA, et al. (2006) Heritable endosymbionts of Drosophila. Genetics 174: 363-376.
2. Werren JH (1997) Biology of Wolbachia. Annu Rev Entomol 42: 587-609.
3. Werren JH, Windsor DM (2000) Wolbachia infection frequencies in insects: evidence of a global equilibrium? Proc Biol Sci 267: 1277-1285.
4. Baldo L, Bordenstein S, Wernegreen JJ, Werren JH (2006) Widespread recombination throughout Wolbachia genomes. Mol Biol Evol 23: 437-449.
5. Werren JH, Baldo L, Clark ME (2008) Wolbachia: master manipulators of invertebrate biology. Nat Rev Microbiol 6: 741-751.
6. Klasson L, Westberg J, Sapountzis P, Naslund K, Lutnaes Y, et al. (2009) The mosaic genome structure of the Wolbachia wRi strain infecting Drosophila simulans. Proc Natl Acad Sci U S A 106: 5725-5730.
7. Teixeira L, Ferreira A, Ashburner M (2008) The bacterial symbiont Wolbachia induces resistance to RNA viral infections in Drosophila melanogaster. PLoS Biol 6: e2.
8. Hedges LM, Brownlie JC, O'Neill SL, Johnson KN (2008) Wolbachia and virus protection in insects. Science 322: 702.
9. Brownlie JC, Cass BN, Riegler M, Witsenburg JJ, Iturbe-Ormaetxe I, et al. (2009) Evidence for metabolic provisioning by a common invertebrate endosymbiont, Wolbachia pipientis, during periods of nutritional stress. PLoS Pathog 5: e1000368.
10. Jeyaprakash A, Hoy MA (2000) Long PCR improves Wolbachia DNA amplification: wsp sequences found in 76% of sixty-three arthropod species. Insect Mol Biol 9: 393-405.
11. Hilgenboecker K, Hammerstein P, Schlattmann P, Telschow A, Werren JH (2008) How many species are infected with Wolbachia?--A statistical analysis of current data. FEMS Microbiol Lett 281: 215-220.
12. Zug R, Hammerstein P (2012) Still a host of hosts for wolbachia: analysis of recent data suggests that 40% of terrestrial arthropod species are infected. PLoS One 7: e38544.
13. O'Neill SL, Giordano R, Colbert AM, Karr TL, Robertson HM (1992) 16S rRNA phylogenetic analysis of the bacterial endosymbionts associated with cytoplasmic incompatibility in insects. Proc Natl Acad Sci U S A 89: 2699-2702.
14. Jiggins FM, von Der Schulenburg JH, Hurst GDD, Majerus ME (2001) Recombination confounds interpretations of Wolbachia evolution. Proc Biol Sci 268: 1423-1427.
15. Kraaijeveld K, Franco P, de Knijff P, Stouthamer R, van Alphen JJ (2011) Clonal genetic variation in a Wolbachia-infected asexual wasp: horizontal transmission or historical sex? Mol Ecol 20: 3644-3652.
16. Hale LR, Hoffmann AA (1990) Mitochondrial DNA Polymorphism and Cytoplasmic Incompatibility in Natural Populations of Drosophila simulans. Evolution 44: 1383-1386.
17. Solignac M, Vautrin D, Rousset F (1994) Widespread occurence of the proteobacteria Wolbachia and partial cytoplasmic incompatibility in Drosophila melanogaster. Comptes rendus de l'Académie des sciences Série 3, Sciences de la vie 317: 461-470.
18. Jiggins FM (2003) Male-killing Wolbachia and mitochondrial DNA: selective sweeps, hybrid introgression and parasite population dynamics. Genetics 164: 5-12.
19. Nunes MD, Nolte V, Schlotterer C (2008) Nonrandom Wolbachia infection status of Drosophila melanogaster strains with different mtDNA haplotypes. Mol Biol Evol 25: 2493-2498.

20. Charlat S, Duplouy A, Hornett EA, Dyson EA, Davies N, et al. (2009) The joint evolutionary histories of Wolbachia and mitochondria in Hypolimnas bolina. BMC Evol Biol 9: 64.
21. Turelli M, Hoffmann AA, McKechnie SW (1992) Dynamics of cytoplasmic incompatibility and mtDNA variation in natural Drosophila simulans populations. Genetics 132: 713-723.
22. Ballard JW, Hatzidakis, J., Karr, T.L., Kreitman, M. (1996) Reduced variation in Drosophila simulans mitochondrial DNA. Genetics 144: 1519-1528.
23. Ballard JW, Kreitman, M. (1994) Unraveling selection in the mitochondrial genome of Drosophila. Genetics 138: 757-772.
24. Ballard JW (2000) Comparative genomics of mitochondrial DNA in members of the Drosophila melanogaster subgroup. J Mol Evol 51: 48-63.
25. Dyer KA, Jaenike J (2004) Evolutionarily stable infection by a male-killing endosymbiont in Drosophila innubila: molecular evidence from the host and parasite genomes. Genetics 168: 1443-1455.
26. Rousset F, Vautrin D, Solignac M (1992) Molecular identification of Wolbachia, the agent of cytoplasmic incompatibility in Drosophila simulans, and variability in relation with host mitochondrial types. Proc Biol Sci 247: 163-168.
27. Holden PR, Jones P, Brookfield JF (1993) Evidence for a Wolbachia symbiont in Drosophila melanogaster. Genet Res 62: 23-29.
28. Wu M, Sun LV, Vamathevan J, Riegler M, Deboy R, et al. (2004) Phylogenomics of the reproductive parasite Wolbachia pipientis wMel: a streamlined genome overrun by mobile genetic elements. PLoS Biol 2: E69.
29. Hoffmann AA (1988) Partial cytoplasmic incompatibility between two Australian populations of Drosophila melanogaster. Entomologia Experimentalis et Applicata 48: 61-67.
30. Hoffmann AA, Clancy DJ, Merton E (1994) Cytoplasmic incompatibility in Australian populations of Drosophila melanogaster. Genetics 136: 993-999.
31. Ilinsky YY, Zakharov IK (2007) The endosymbiont Wolbachia in Eurasian populations of Drosophila melanogaster. Genetika 43: 905-915.
32. Verspoor RL, Haddrill PR (2011) Genetic diversity, population structure and Wolbachia infection status in a worldwide sample of Drosophila melanogaster and D. simulans populations. PLoS One 6: e26318.
33. Bourtzis K, Nirgianaki A, Onyango P, Savakis C (1994) A prokaryotic dnaA sequence in Drosophila melanogaster: Wolbachia infection and cytoplasmic incompatibility among laboratory strains. Insect Mol Biol 3: 131-142.
34. Zhou W, Rousset F, O'Neil S (1998) Phylogeny and PCR-based classification of Wolbachia strains using wsp gene sequences. Proc Biol Sci 265: 509-515.
35. Riegler M, Sidhu M, Miller WJ, O'Neill SL (2005) Evidence for a global Wolbachia replacement in Drosophila melanogaster. Curr Biol 15: 1428-1433.
36. Salzberg SL, Dunning Hotopp JC, Delcher AL, Pop M, Smith DR, et al. (2005) Serendipitous discovery of Wolbachia genomes in multiple Drosophila species. Genome Biol 6: R23.
37. Montooth KL, Abt DN, Hofmann JW, Rand DM (2009) Comparative genomics of Drosophila mtDNA: Novel features of conservation and change across functional domains and lineages. J Mol Evol 69: 94-114.
38. Mackay TFC, Richards S, Stone EA, Barbadilla A, Ayroles JF, et al. (2012) The Drosophila melanogaster Genetic Reference Panel. Nature 482: 173-178.

39. Pool JE, Corbett-Detig RB, Sugino RP, Stevens KA, Cardeno CM, et al. (submitted) Population genomics of sub-Saharan Drosophila melanogaster: African diversity and non-African admixture.
40. Dunning Hotopp JC, Clark ME, Oliveira DC, Foster JM, Fischer P, et al. (2007) Widespread lateral gene transfer from intracellular bacteria to multicellular eukaryotes. Science 317: 1753-1756.
41. Boyle L, O'Neill SL, Robertson HM, Karr TL (1993) Interspecific and intraspecific horizontal transfer of Wolbachia in Drosophila. Science 260: 1796-1799.
42. Hurst GDD, Jiggins FM (2005) Problems with mitochondrial DNA as a marker in population, phylogeographic and phylogenetic studies: the effects of inherited symbionts. Proc Biol Sci 272: 1525-1534.
43. Hudson RR, Boos DD, Kaplan NL (1992) A statistical test for detecting geographic subdivision. Mol Biol Evol 9: 138-151.
44. Kondo R, Satta Y, Matsuura ET, Ishiwa H, Takahata N, et al. (1990) Incomplete maternal transmission of mitochondrial DNA in Drosophila. Genetics 126: 657-663.
45. Hoffmann AA, Turelli M (1988) Unidirectional incompatibility in Drosophila simulans: inheritance, geographic variation and fitness effects. Genetics 119: 435-444.
46. Baele G, Lemey P, Bedford T, Rambaut A, Suchard MA, et al. (2012) Improving the accuracy of demographic and molecular clock model comparison while accommodating phylogenetic uncertainty. Mol Biol Evol 29: 2157-2167.
47. Keightley PD, Trivedi U, Thomson M, Oliver F, Kumar S, et al. (2009) Analysis of the genome sequences of three Drosophila melanogaster spontaneous mutation accumulation lines. Genome Res 19: 1195-1201.
48. Haag-Liautard C, Coffey N, Houle D, Lynch M, Charlesworth B, et al. (2008) Direct estimation of the mitochondrial DNA mutation rate in Drosophila melanogaster. PLoS Biol 6: e204.
49. Li H, Stephan W (2006) Inferring the demographic history and rate of adaptive substitution in Drosophila. PLoS Genet 2: e166.
50. Haag-Liautard C, Dorris M, Maside X, Macaskill S, Halligan DL, et al. (2007) Direct estimation of per nucleotide and genomic deleterious mutation rates in Drosophila. Nature 445: 82-85.
51. Stephan W, Li H (2007) The recent demographic and adaptive history of *Drosophila melanogaster.* Heredity 98: 65-68.
52. David JR, Capy P (1988) Genetic variation of Drosophila melanogaster natural populations. Trends Genet 4: 106-111.
53. Keller A (2007) Drosophila melanogaster's history as a human commensal. Curr Biol 17: R77-81.
54. Ilinsky YY, Zakharov IK (2006) Genetic correlation between types of mtDNA of Drosophila melanogaster and genotypes of its primary endosymbiont, Wolbachia. Dros Inf Serv 89: 89-90.
55. Linheiro RS, Bergman CM (2012) Whole Genome Resequencing Reveals Natural Target Site Preferences of Transposable Elements in Drosophila melanogaster. PLoS One 7: e30008.
56. Pirrotta V, Brockl C (1984) Transcription of the Drosophila white locus and some of its mutants. EMBO J 3: 563-568.
57. Iturbe-Ormaetxe I, Woolfit M, Rances E, Duplouy A, O'Neill SL (2011) A simple protocol to obtain highly pure Wolbachia endosymbiont DNA for genome sequencing. J Microbiol Methods 84: 134-136.

58. Lewis DL, Farr CL, Kaguni LS (1995) Drosophila melanogaster mitochondrial DNA: completion of the nucleotide sequence and evolutionary comparisons. Insect Mol Biol 4: 263-278.
59. Garesse R (1988) Drosophila melanogaster mitochondrial DNA: gene organization and evolutionary considerations. Genetics 118: 649-663.
60. de Bruijn MH (1983) Drosophila melanogaster mitochondrial DNA, a novel organization and genetic code. Nature 304: 234-241.
61. Rand DM, Dorfsman M, Kann LM (1994) Neutral and non-neutral evolution of Drosophila mitochondrial DNA. Genetics 138: 741-756.
62. Nunes MD, Neumeier H, Schlotterer C (2008) Contrasting patterns of natural variation in global Drosophila melanogaster populations. Mol Ecol 17: 4470-4479.
63. James AC, Ballard JW (2003) Mitochondrial genotype affects fitness in Drosophila simulans. Genetics 164: 187-194.
64. Weeks AR, Turelli M, Harcombe WR, Reynolds KT, Hoffmann AA (2007) From parasite to mutualist: rapid evolution of Wolbachia in natural populations of Drosophila. PLoS Biol 5: e114.
65. Kumar S, Blaxter ML (2012) Simultaneous genome sequencing of symbionts and their hosts. Symbiosis 55: 119-126.
66. Langley CH, Crepeau M, Cardeno C, Corbett-Detig R, Stevens K (2011) Circumventing heterozygosity: sequencing the amplified genome of a single haploid Drosophila melanogaster embryo. Genetics 188: 239-246.
67. Correa CC, Aw WC, Melvin RG, Pichaud N, Ballard JW (2012) Mitochondrial DNA variants influence mitochondrial bioenergetics in Drosophila melanogaster. Mitochondrion 12: 459-464.
68. Hoffmann AA, Hercus M, Dagher H (1998) Population dynamics of the Wolbachia infection causing cytoplasmic incompatibility in Drosophila melanogaster. Genetics 148: 221-231.
69. Hoffmann AA, Turelli M (1997) Cytoplasmic incompatibility in insects. In: O'Neil SL, Hoffmann AA, Werren JH, editors. Influential Passengers: Inherited Microorganisms and Invertebrate Reproduction. Oxford: Oxford University Press. pp. 42-80.
70. Funk DJ, Helbling L, Wernegreen JJ, Moran NA (2000) Intraspecific phylogenetic congruence among multiple symbiont genomes. Proc Biol Sci 267: 2517-2521.
71. Nigro L, Prout T (1990) Is there selection on RFLP differences in mitochondrial DNA? Genetics 125: 551-555.
72. Baldo L, Ayoub NA, Hayashi CY, Russell JA, Stahlhut JK, et al. (2008) Insight into the routes of Wolbachia invasion: high levels of horizontal transfer in the spider genus Agelenopsis revealed by Wolbachia strain and mitochondrial DNA diversity. Mol Ecol 17: 557-569.
73. Moran N, Munson M, Baumann P, Ishikawa H (1993) A Molecular Clock in Endosymbiotic Bacteria is Calibrated Using the Insect Hosts. Proceedings of the Royal Society of London Series B: Biological Sciences 253: 167-171.
74. Ochman H, Elwyn S, Moran NA (1999) Calibrating bacterial evolution. Proc Natl Acad Sci U S A 96: 12638-12643.
75. Rocha EP, Smith JM, Hurst LD, Holden MT, Cooper JE, et al. (2006) Comparisons of dN/dS are time dependent for closely related bacterial genomes. J Theor Biol 239: 226-235.
76. Moran NA, McLaughlin HJ, Sorek R (2009) The dynamics and time scale of ongoing genomic erosion in symbiotic bacteria. Science 323: 379-382.
77. Clark MA, Moran NA, Baumann P (1999) Sequence evolution in bacterial endosymbionts having extreme base compositions. Mol Biol Evol 16: 1586-1598.

78. Moran NA, Mira A (2001) The process of genome shrinkage in the obligate symbiont Buchnera aphidicola. Genome Biol 2: RESEARCH0054.
79. Min KT, Benzer S (1997) Wolbachia, normally a symbiont of Drosophila, can be virulent, causing degeneration and early death. Proc Natl Acad Sci U S A 94: 10792-10796.
80. Chandler JA, Lang JM, Bhatnagar S, Eisen JA, Kopp A (2011) Bacterial communities of diverse Drosophila species: ecological context of a host-microbe model system. PLoS Genet 7: e1002272.
81. Pool JE, Wong A, Aquadro CF (2006) Finding of male-killing Spiroplasma infecting Drosophila melanogaster in Africa implies transatlantic migration of this endosymbiont. Heredity (Edinb) 97: 27-32.
82. Ventura IM, Martins AB, Lyra ML, Andrade CA, Carvalho KA, et al. (2012) Spiroplasma in Drosophila melanogaster Populations: Prevalence, Male-Killing, Molecular Identification, and No Association with Wolbachia. Microb Ecol 64: 794-801.
83. Li H, Durbin R (2009) Fast and accurate short read alignment with Burrows-Wheeler transform. Bioinformatics 25: 1754-1760.
84. Li H, Handsaker B, Wysoker A, Fennell T, Ruan J, et al. (2009) The Sequence Alignment/Map format and SAMtools. Bioinformatics 25: 2078-2079.
85. Brizuela BJ, Elfring L, Ballard J, Tamkun JW, Kennison JA (1994) Genetic analysis of the brahma gene of Drosophila melanogaster and polytene chromosome subdivisions 72AB. Genetics 137: 803-813.
86. Quinlan AR, Hall IM (2010) BEDTools: a flexible suite of utilities for comparing genomic features. Bioinformatics 26: 841-842.
87. Stajich JE, Block D, Boulez K, Brenner SE, Chervitz SA, et al. (2002) The Bioperl toolkit: Perl modules for the life sciences. Genome Res 12: 1611-1618.
88. Jiggins FM, Tinsley MC (2005) An ancient mitochondrial polymorphism in Adalis bipunctata linked to a sex-ratio-distorting bacterium. Genetics 171: 1115-1124.
89. Magwire MM, Bayer F, Webster CL, Cao C, Jiggins FM (2011) Successive increases in the resistance of Drosophila to viral infection through a transposon insertion followed by a Duplication. PLoS Genet 7: e1002337.
90. Stamatakis A (2006) RAxML-VI-HPC: maximum likelihood-based phylogenetic analyses with thousands of taxa and mixed models. Bioinformatics 22: 2688-2690.
91. Stamatakis A, Hoover P, Rougemont J (2008) A rapid bootstrap algorithm for the RAxML Web servers. Syst Biol 57: 758-771.
92. Bingham J, Sudarsanam S (2000) Visualizing large hierarchical clusters in hyperbolic space. Bioinformatics 16: 660-661.
93. Librado P, Rozas J (2009) DnaSP v5: a software for comprehensive analysis of DNA polymorphism data. Bioinformatics 25: 1451-1452.
94. Drummond AJ, Rambaut A (2007) BEAST: Bayesian evolutionary analysis by sampling trees. BMC Evol Biol 7: 214.
95. Drummond AJ, Suchard MA, Xie D, Rambaut A (2012) Bayesian phylogenetics with BEAUti and the BEAST 1.7. Mol Biol Evol 29: 1969-1973.
96. Lemey P, Rambaut A, Drummond AJ, Suchard MA (2009) Bayesian phylogeography finds its roots. PLoS Comput Biol 5: e1000520.
97. Edwards CJ, Suchard MA, Lemey P, Welch JJ, Barnes I, et al. (2011) Ancient hybridization and an Irish origin for the modern polar bear matriline. Curr Biol 21: 1251-1258.
98. Weinert LA, Welch JJ, Suchard MA, Lemey P, Rambaut A, et al. (2012) Molecular dating of human-to-bovid host jumps by Staphylococcus aureus reveals an association with the spread of domestication. Biol Lett 8: 829-832.

99. Shapiro B, Rambaut A, Drummond AJ (2006) Choosing appropriate substitution models for the phylogenetic analysis of protein-coding sequences. Mol Biol Evol 23: 7-9.
100. Shah SD, Doorbar J, Goldstein RA (2010) Analysis of host-parasite incongruence in papillomavirus evolution using importance sampling. Mol Biol Evol 27: 1301-1314.
101. Nei M, Li WH (1979) Mathematical model for studying genetic variation in terms of restriction endonucleases. Proc Natl Acad Sci U S A 76: 5269-5273.
102. Watterson GA (1975) On the number of segregating sites in genetical models without recombination. Theor Popul Biol 7: 256-276.
103. Hutter S, Vilella AJ, Rozas J (2006) Genome-wide DNA polymorphism analyses using VariScan. BMC Bioinformatics 7: 409.
104. Tajima F (1989) Statistical method for testing the neutral mutation hypothesis by DNA polymorphism. Genetics 123: 585-595.
105. Hudson RR (2002) Generating samples under a Wright-Fisher neutral model of genetic variation. Bioinformatics 18: 337-338.
106. Wall JD, Hudson RR (2001) Coalescent simulations and statistical tests of neutrality. Mol Biol Evol 18: 1134-1135; author reply 1136-1138.

**Figures Legends**

**Figure 1**: Relationship between depth and breadth of sequencing coverage for *Wolbachia* and mtDNA assemblies in the DGRP (A) and DPGP (B) samples. Depth of coverage is calculated as the number of reads present at each nucleotide in the reference sequence averaged over all sites and is show in $\log_{10}$ units. Breadth of coverage is defined as the proportion of nucleotides in the consensus sequence (excluding ambiguous characters and indels) relative to the length of the reference sequence. Results of assays for *Wolbachia* infection using PCR are labelled for the DGRP population. Strains with a depth >1 and a breadth >90% were defined as infected by *in silico* criteria.

**Figure 2**: Relative depth of sequencing coverage for *Wolbachia* and mtDNA assemblies. Mean depth of coverage across the entire *Wolbachia* or mtDNA assembly was normalized relative to mean depth of coverage of a nuclear locus from the *D. melanogaster* genome on chromosome 3L of equivalent size to the *Wolbachia* genome. Normalized coverage is shown separately for mtDNA from infected and uninfected to allow direct comparison to the same set of strains infected with *Wolbachia*.

**Figure 3**: Maximum likelihood genealogy of the *D. melanogaster* mtDNA in the DGRP and DPGP strains. Strains infected with *Wolbachia* are shows in black font and are labelled with asterisks, while uninfected strains are shown in grey font. Strains highlighted with a grey background are from the North American DGRP sample, underlined strains are from the European DPGP sample, and the remainder are from the African DPGP sample. The dm3 and NC_001709 reference sequences are found in clades III and VI, respectively. The numerical suffix after strain names represents the COI haplotype according to Nunes *et al.* [19] with numbering extended to new haplotypes discovered here. The underlying data consist of an ungapped multiple alignment of 292 sequences each of 12,225 bp in length. The unrooted ML tree was midpoint rooted for visualization and branches with >85% RAxML bootstrap values are shown with coloured boxes. Scale bars for branch lengths are in terms of mutations per site.

**Figure 4**: Maximum likelihood genealogies of mtDNA (A) and *Wolbachia* (B) from infected DGRP and DPGP strains. Strains highlighted with a grey background are from the North American DGRP sample and the remainder are from the African and European DPGP sample. The major cytoplasmic lineages discussed in the main text are shown as clades I-V. Reference

sequences for mtDNA (dm3 and NC_001079) and *Wolbachia* (AE017196) are labelled and dashed arrows represent their positions on the tree. Asterisks represent two Ugandan strains that are also predicted to be infected with *Spiroplasma*. The mitochondrial tree is based on an ungapped multiple alignment of 181 sequences of 12,236 bp in length, and the *Wolbachia* tree is based on an ungapped multiple alignment of 180 sequences of 957,546 bp in length. Unrooted ML trees were midpoint rooted for visualization and branches with >80% RAxML bootstrap values are shown with coloured boxes. Scale bars for branch lengths are in terms of mutations per site.

**Figure 5**: Schematic representation of dated Bayesian genealogy of *Wolbachia* in *D. melanogaster*. Dates (in years) are median estimates with 95% Bayesian Credible Intervals. The major cytoplasmic lineages discussed in the main text are shown as clades I-VI and VI. Clade V is only comprised of uninfected strains in this sample and not shown here.

**Tables**

**Table 1:** Infection status of *Wolbachia* in the DGRP and DPGP samples.

| Sample | Total Number of Strains | Number of Infected Strains | % Infected | 95% Confidence Interval * |
|---|---|---|---|---|
| DGRP+DPGP | 290 | 179 | 61.8 | 0.56-0.68 |
| DGRP | 174 | 91 | 52.2 | 0.45-0.60 |
| DPGP | 116 | 88 | 75.9 | 0.67-0.83 |

* Based on the assumption of a binomial distribution.

**Table 2:** Bayesian estimates of evolutionary rates in *Wolbachia* and mitochondrial genomes. $L$ is the number of sites analyzed, and $S$ is the number of variable sites. Evolutionary rates are in terms of *D. melanogaster* generations.

| Site class | $L$ | $S$ | Median rate (substitutions/ site/generation) | 95% Credible Interval |
|---|---|---|---|---|
| *Wolbachia* 1$^{st}$ + 2$^{nd}$ codon positions | 517,885 | 345 | $6.42 \times 10^{-10}$ | $2.76 \times 10^{-10}$ - $1.21 \times 10^{-9}$ |
| *Wolbachia* 3$^{rd}$ codon positions | 259,051 | 195 | $6.87 \times 10^{-10}$ | $2.88 \times 10^{-10}$ - $1.29 \times 10^{-9}$ |
| *Wolbachia* noncoding DNA positions | 178,739 | 180 | $8.00 \times 10^{-10}$ | $3.28 \times 10^{-10}$ - $1.50 \times 10^{-9}$ |
| mtDNA 1$^{st}$ + 2$^{nd}$ codon positions | 6,608 | 86 | $1.36 \times 10^{-8}$ | $5.58 \times 10^{-9}$ - $2.64 \times 10^{-8}$ |
| mtDNA 3$^{rd}$ codon positions * | 3,300 | 149 | $6.89 \times 10^{-8}$ | $3.00 \times 10^{-8}$ - $1.24 \times 10^{-7}$ |

* Set as a prior using data from [48]

**Table 3:** Summary of polymorphism in *Wolbachia* and mitochondrial genomes in the DGRP sample. *N* is the number of strains, *L* is the number of sites analyzed, and *S* is the number of variable sites. Estimates of π and θ (per site) are based on the number of mutations. All Tajima's *D* tests had *P*<0.01 based on 10,000 coalescent simulations that assumed no recombination and were conditioned on the number of variable sites.

| Genome | N | L | S | π | θ | Tajima's D |
|---|---|---|---|---|---|---|
| *Wolbachia* (Infected strains) | 91 | 1,209,286 | 205 | 0.0000106 | 0.0000334 | -2.31443 |
| mtDNA (Infected strains) | 91 | 14,492 | 115 | 0.0005337 | 0.0015749 | -2.20772 |
| mtDNA (All strains) | 174 | 14,408 | 164 | 0.0004381 | 0.0019974 | -2.48508 |

# Supporting Information

**Dataset S1**: Summary of metadata, infection status, input sequence data, assembly properties and phylogenetic classification for each strain in the DGRP and DPGP datasets.

**Dataset S2**: Archive of maximum likelihood phylogenetic trees in Newick format based on alignments of the *Wolbachia* and mitochondrial genomes. Branch lengths are in substitutions per site and internal node labels represent bootstrap probabilities based on the RAxML rapid bootstrap algorithm.

**Dataset S3**: Archive of Bayesian dated phylogenetic trees in Newick format and corresponding BEAST XML files based on a concatenated alignment of the *Wolbachia* and mitochondrial genomes.

**Dataset S4**: Tables of variable sites for *Wolbachia* and mitochondrial genomes. Variant sites were extracted from the alignments used for BEAST analysis (indels and fully ambiguous sites removed, no reference sequences, with annotation track). Coordinates are provided relative to reference genome (ref) and relative to the alignment with columns containing indels and fully ambiguous sites removed (aln). Classes of annotated sites (ann) are defined in the Materials and Methods.

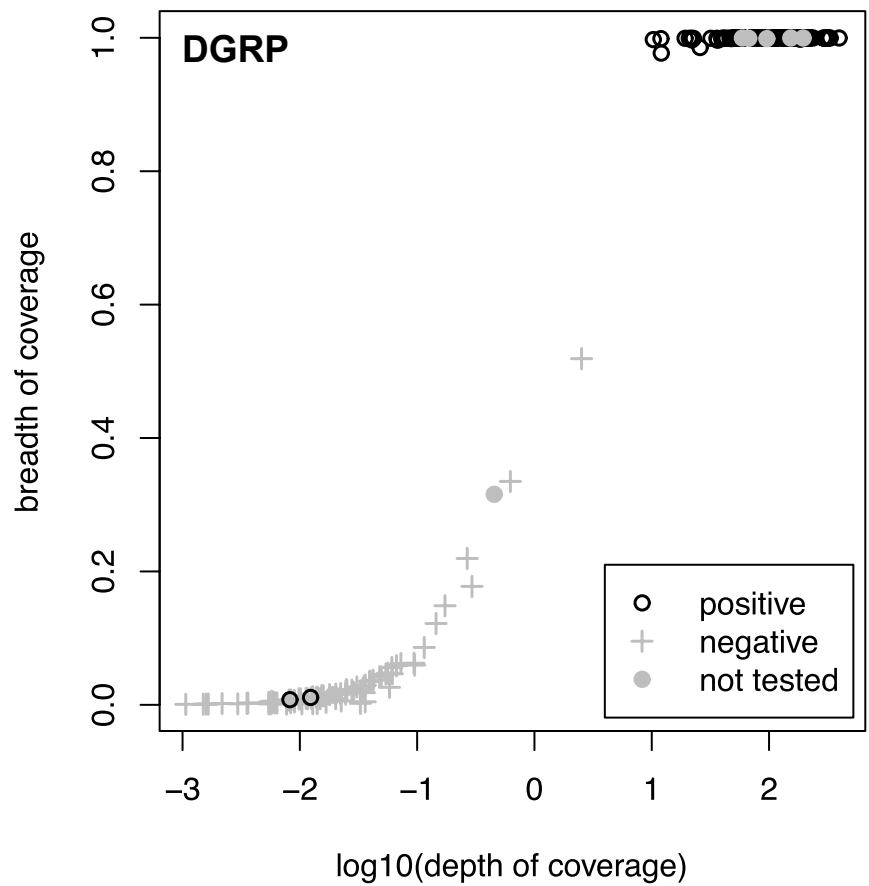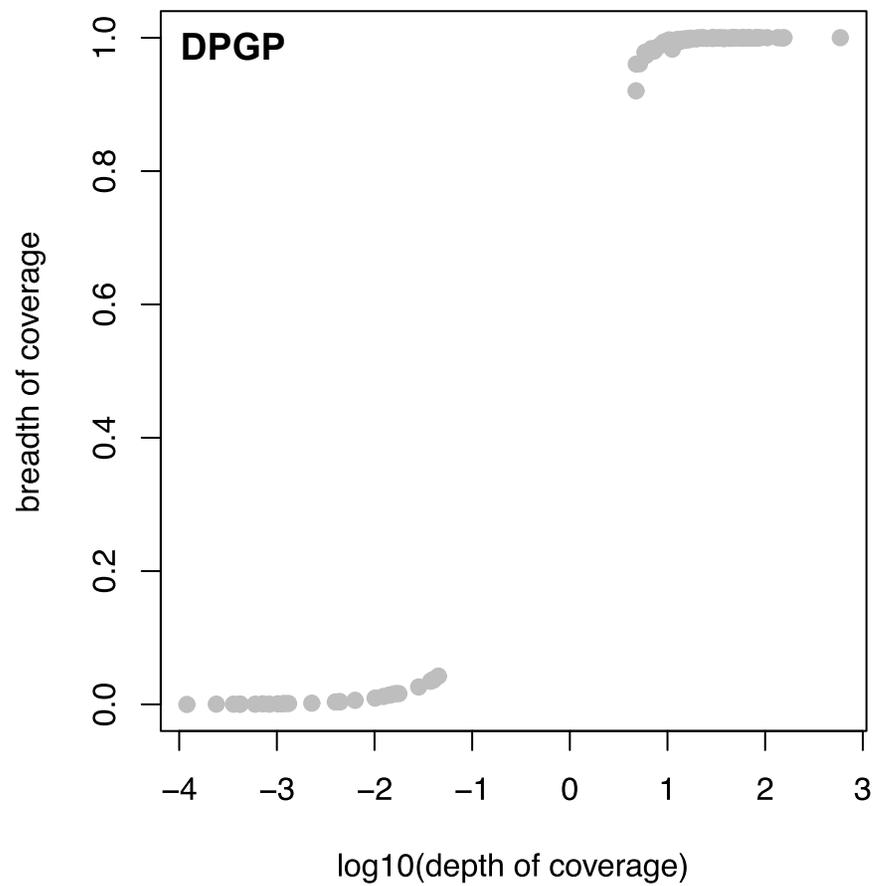

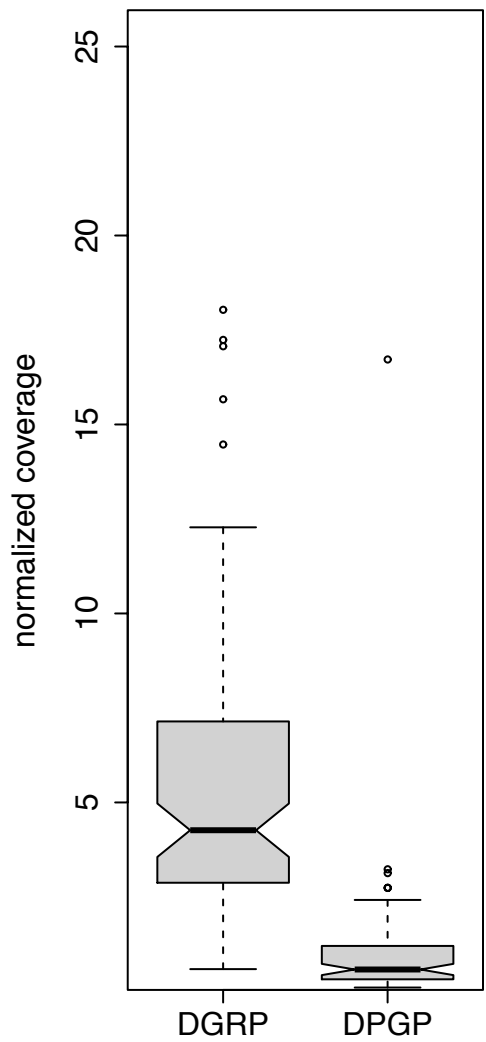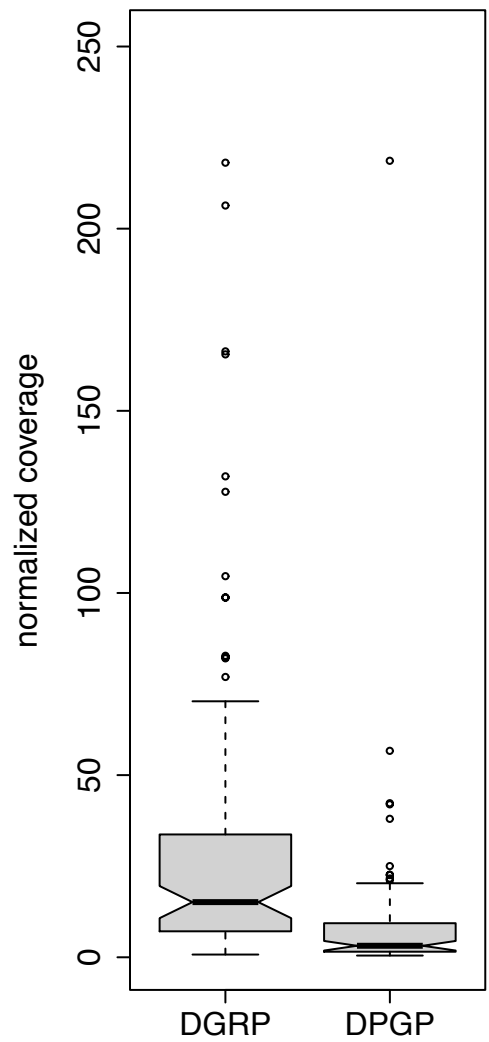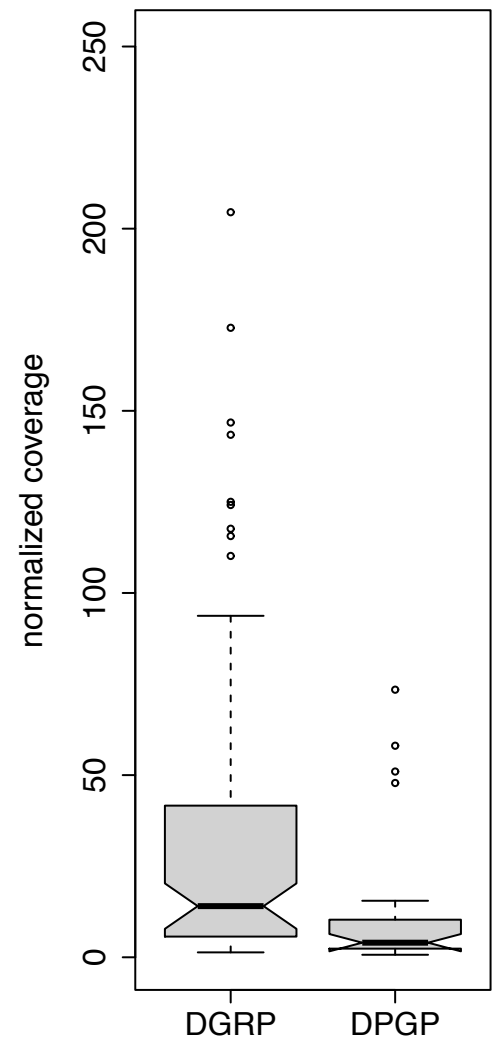

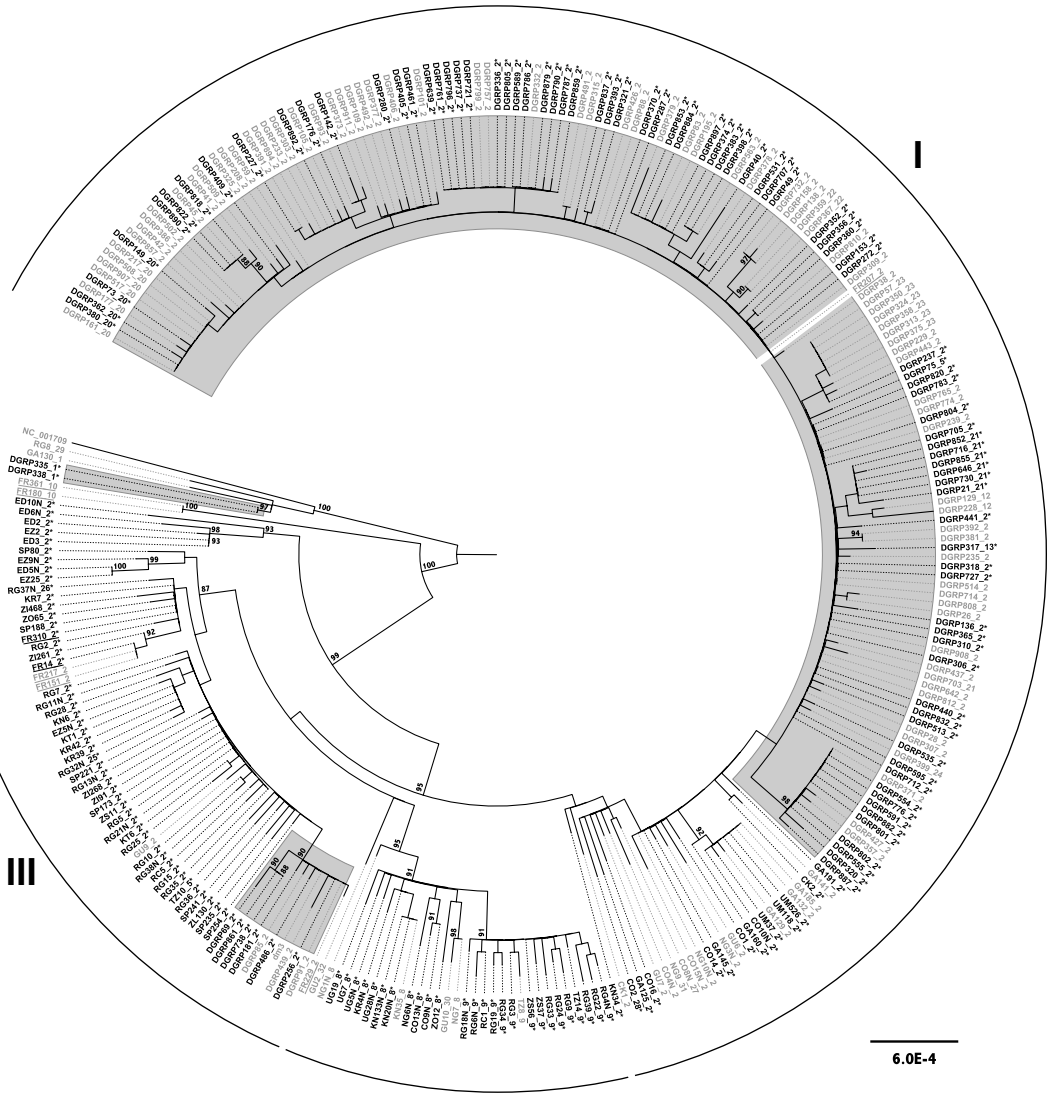

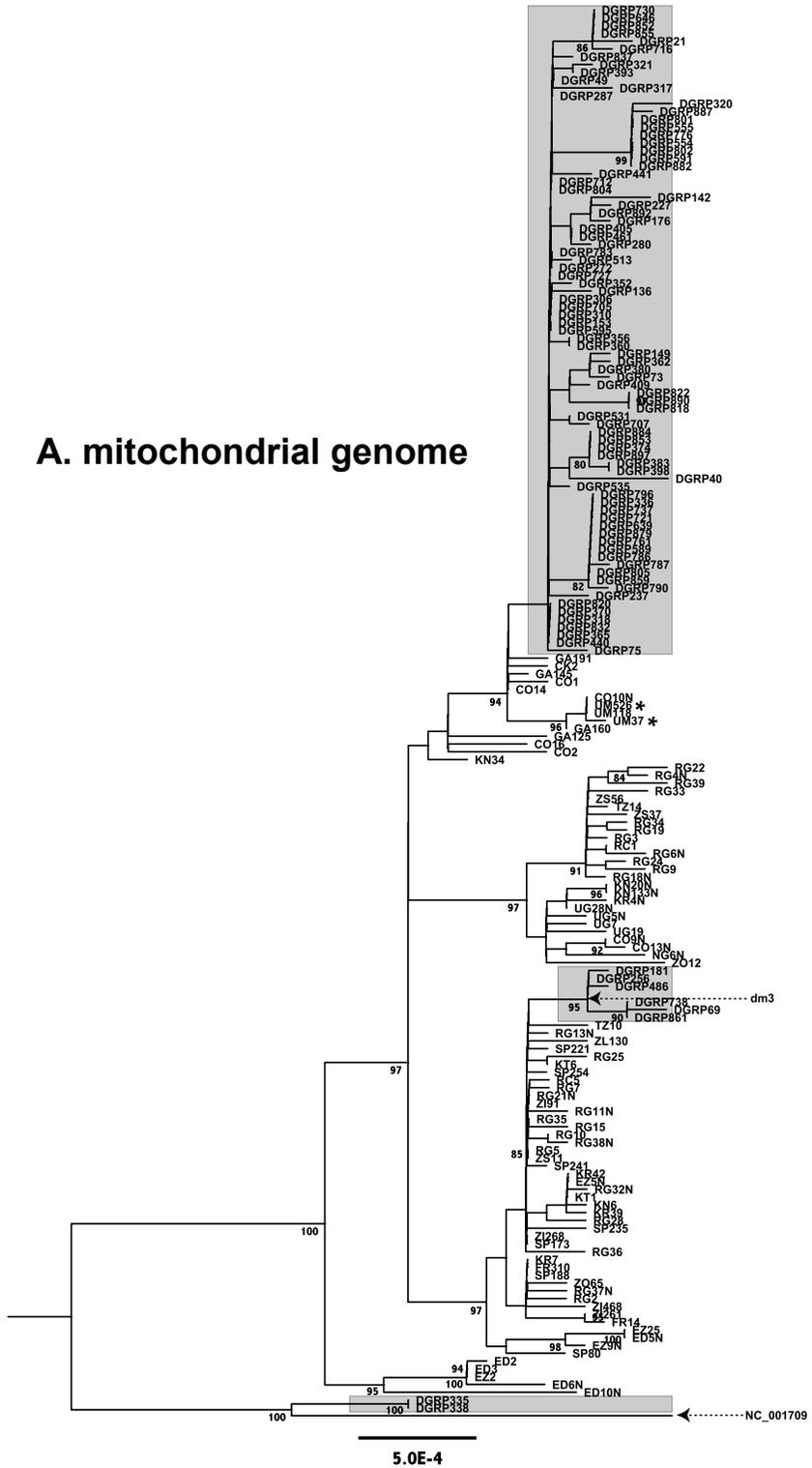
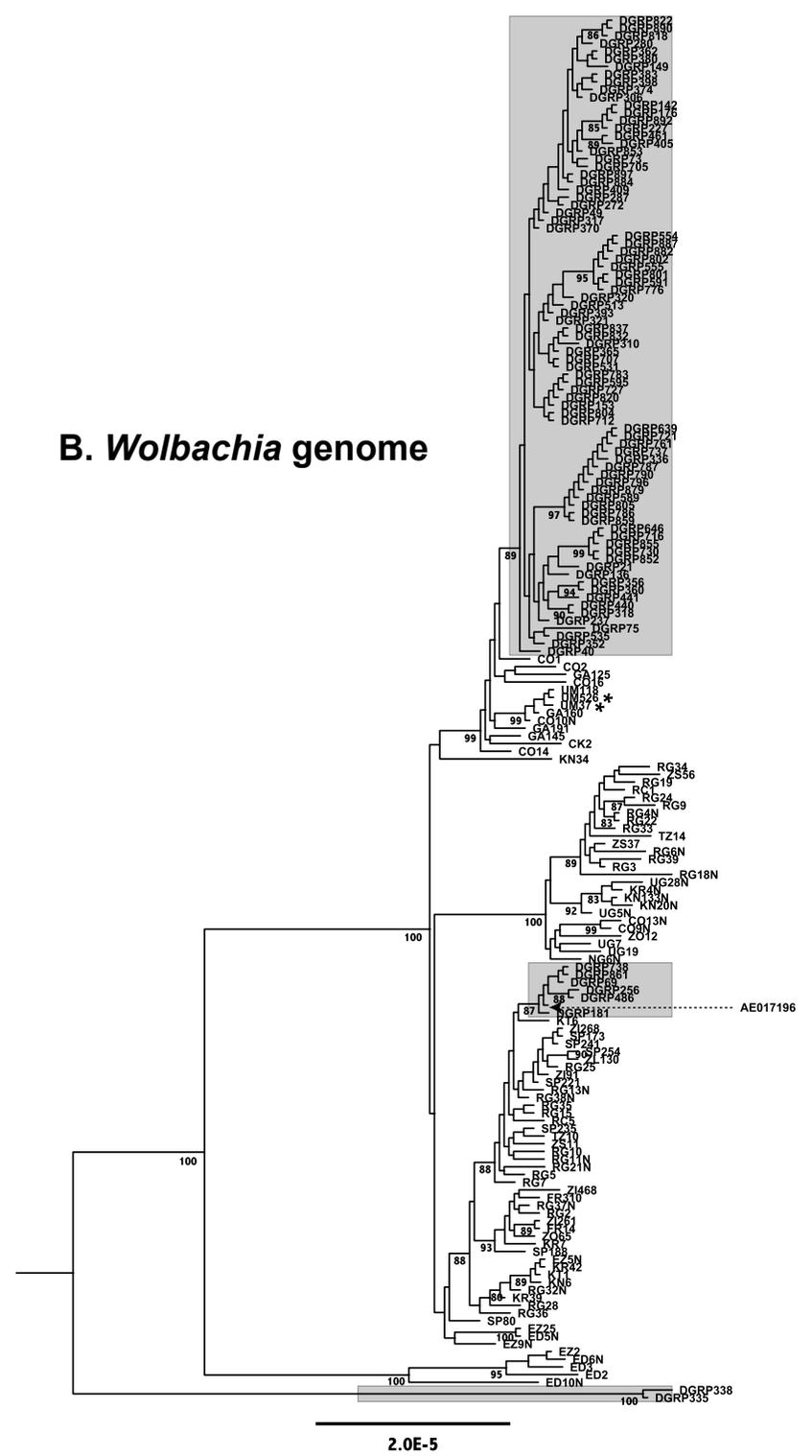

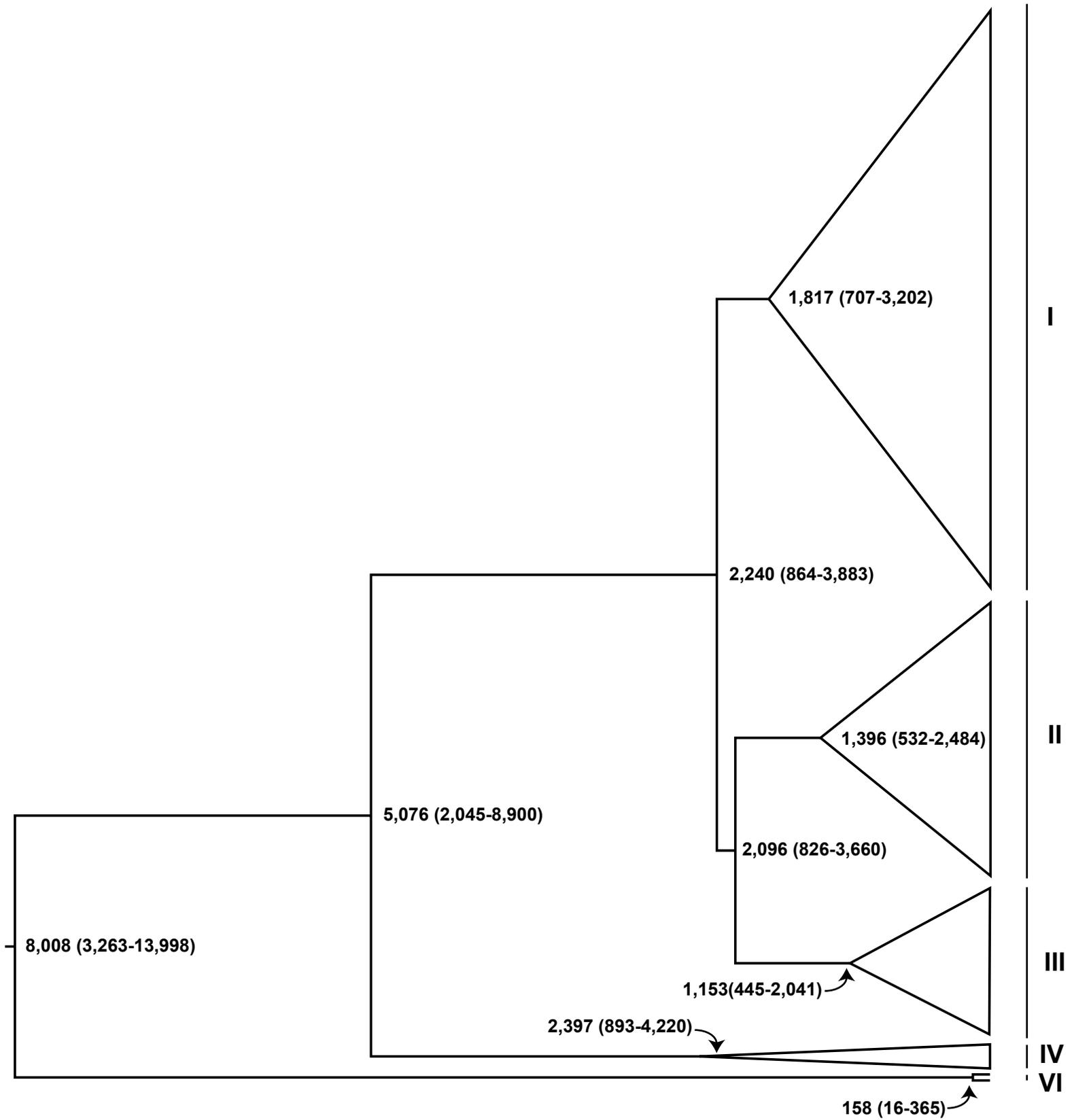